\documentclass[aps,prd,superscriptaddress,nofootinbib,amsmath,amsfonts,preprintnumbers,groupedaddress,showpacs,10pt,english,dvi2pdfmx]{revtex4-1}
\usepackage{amsmath}
\usepackage{amssymb}
\usepackage{babel}  
\usepackage{wrapfig}
\usepackage{cancel}
\newcommand{\nn}{\nonumber \\}
\newcommand{\e}{\mathrm{e}}
\usepackage{relsize,exscale}
\makeatletter

\usepackage{array,multirow,graphicx}
\usepackage{dcolumn}
\usepackage{newlfont}
\usepackage{bm}
\usepackage[colorlinks,citecolor=blue,urlcolor=blue,linkcolor=blue]{hyperref}
\usepackage[figtopcap]{subfigure}
\usepackage{color}
\usepackage{verbatim}

\newcommand{\la}\lambda

\renewcommand{\k}\kappa

\def\be{\begin{align}}
\def\ee{\end{align}}
\def\bea{\begin{eqnarray}}
\def\eea{\end{eqnarray}}
\def\bal{\begin{align}}
\def\eal{\end{align}}

\usepackage{scalerel}
\usepackage{tikz}
\usetikzlibrary{svg.path}
\definecolor{orcidlogocol}{HTML}{A6CE39}
\tikzset{
 orcidlogo/.pic={
 \fill[orcidlogocol] svg{M256,128c0,70.7-57.3,128-128,128C57.3,256,0,198.7,0,128C0,57.3,57.3,0,128,0C198.7,0,256,57.3,256,128z};
 \fill[white] svg{M86.3,186.2H70.9V79.1h15.4v48.4V186.2z}
 svg{M108.9,79.1h41.6c39.6,0,57,28.3,57,53.6c0,27.5-21.5,53.6-56.8,53.6h-41.8V79.1z M124.3,172.4h24.5c34.9,0,42.9-26.5,42.9-39.7c0-21.5-13.7-39.7-43.7-39.7h-23.7V172.4z}
 svg{M88.7,56.8c0,5.5-4.5,10.1-10.1,10.1c-5.6,0-10.1-4.6-10.1-10.1c0-5.6,4.5-10.1,10.1-10.1C84.2,46.7,88.7,51.3,88.7,56.8z};}}
\newcommand\orcid[1]{\href{https://orcid.org/#1}{\mbox{\scalerel*{
\begin{tikzpicture}[yscale=-1,transform shape]
\pic{orcidlogo};
\end{tikzpicture}
}{|}}}}
\graphicspath{{./}{Figs/}}

\allowdisplaybreaks[4]

\begin{document}

\tolerance=5000

\date{\today}
\title{Stable gravastar with large surface redshift in Einstein's gravity with two scalar fields}
\author{Shin'ichi~Nojiri~\orcid{0000-0002-0773-8011}}
\email{nojiri@gravity.phys.nagoya-u.ac.jp}
\affiliation{Department of Physics, Nagoya University, Nagoya 464-8602,
Japan \\
\& \\
Kobayashi-Maskawa Institute for the Origin of Particles and the Universe,
Nagoya University, Nagoya 464-8602, Japan }
\author{G.~G.~L.~Nashed~\orcid{0000-0001-5544-1119}}
\email{nashed@bue.edu.eg}
\affiliation {Centre for Theoretical Physics, The British University in Egypt, P.O. Box
43, El Sherouk City, Cairo 11837, Egypt}

\begin{abstract}
We propose a class of models, in which stable gravastar with large surface redshift becomes a solution.
In recent decades, gravastars have become a plausible substitute for black holes.
Researchers have explored stable gravastar models in various alternative gravity theories, in addition to the conventional framework of general relativity.
In this paper, we present a stellar model within the framework of Einstein's gravity with two scalar fields, in accordance with the conjecture proposed
by Mazur and Mottola [Proc. Nat. Acad. Sci. \textbf{101} (2004), 9545-9550].
In the model, the two scalar fields do not propagate by imposing constraints in order to avoid ghosts.
The gravastar comprises two distinct regions, namely: (a) the interior region and (b) the exterior region.
We assume the interior region consists of the de Sitter spacetime, and the exterior region is the Schwarzschild one.
The two regions are connected with each other by the shell region.
On the shell, we assume that the metric is given by a polynomial function of the radial coordinate $r$.
The function has six constants.
These constants are fixed by the smooth junction conditions, i.e., the interior region with the interior layer of the shell and the exterior region with the exterior layer of the shell.
 From these boundary conditions, we are able to write the coefficients of the scalar fields in terms of the interior radius and exterior radius.
To clarify the philosophy of this study, we also give two examples of spacetimes that asymptote as the de Sitter spacetime for small $r$ and
as the Schwarzschild spacetime for large $r$.
Exploration is focused on the physical attribute of the shell region, specifically, its proper length.
The gravastar model's stability has frequently been examined by analyzing the relationship between surface redshift and shell thickness, a comparison we also undertake with previous models.
Especially, we show that there exists a stable gravastar with a large surface redshift prohibited by the instability in the previous works.
Furthermore, by checking the effective equation of state parameters, we show that the gravastar geometry realized in this paper by using two scalar fields
could be difficult to generate with ordinary fluid.

\end{abstract}

\maketitle

\newpage

\section{Introduction}

The final stage in the evolutionary journey of a massive celestial object consistently captures the interest of astrophysicists.
A conjecture proposes that during the final phase of stellar evolution when the gravitational mass of the remaining celestial body,
following the creation of a planetary nebula or the occurrence of a supernova explosion that surpasses three solar masses,
the gravitational collapse of the star leads to the creation of an exceptionally dense entity, specifically, a black hole.
Schwarzschild derived the most straightforward black hole solution from Einstein's vacuum field equation in 1916, which is regarded as a significant solution
for a black hole that is static and uncharged.
However, the solution of Schwarzschild exhibits certain limitations attributed to (I) the existence of a singularity within the confines of a black hole, along with
(II) The presence of an event horizon presents substantial, unresolved challenges.
Mazur and Mottola~\cite{Mitra:2002dh, Mazur:2004fk}
were the first to introduce a novel concept for the collapse of stellar objects, incorporating an expanded notion of the Bose-Einstein Condensation
within the gravitational system.
This model proposed the potential creation of a cold, faint, and compact entity, which they named a gravitational vacuum condensed star or gravastar.
The representation of the gravastar must meet all the theoretical prerequisites to serve as a stable concluding phase in stellar evolution
and offer resolutions to the problems linked to conventional black holes.
Following the triumphant detection of gravitational waves (GWs) in 2015 \cite{LIGOScientific:2016aoc}, there has been a presumption
that these GWs emanate from the collision of two substantial black holes.
Given that the observed signs do not conclusively address the fundamental problems linked to black holes, an alternative method is necessary to tackle these limitations.
In such a framework, the gravastar might play a pivotal role in elucidating the eventual stage of stellar evolution.
Although significant observational proof endorsing the existence of gravastars is absent, it remains crucial to explore the notion of gravastars
as a viable alternative for addressing the conceptual dilemmas linked to understanding black holes.

The gravastar is composed of three separate regions: the inner region, a thin intermediate shell, and an outer region.
After the suggestion made by Mazur and Mottola~\cite{Mitra:2002dh, Mazur:2004fk}, it is postulated that the gravastar's inner region exists
in a phase resembling a de Sitter condensate and is believed to be filled with vacuum energy.
Concurrently, the outer region is in a state of complete vacuum known as Schwarzschild vacuum, and these two regions are divided
by an ultra-relativistic matter-thin shell characterized by an exceptionally high density.
The barotropic equation of state (EoS) is defined as $p = \rho$, with $\rho$ representing the matter-energy density and $p$ standing for pressure.
Within a gravastar, the equation of state (EoS) parameter $\omega\equiv \frac{p}{\rho}$ varies across various regions as depicted below: \\
a) In the interior region $(0 \leq r < r_1)$, the EoS parameter is $\omega = -1$, which corresponds to $p = -\rho$, where $\rho$ is density and $p$ represents pressure. \\
b) In the thin shell $(r_1 <r < r_2)$, the EoS parameter is $\omega = 1$, which corresponds to $p =\rho$.\\
c) In the exterior region $r_2<r$ and $\omega=0$ which corresponds to $p =0$.\\
In this context, $r_1$ and $r_2$ represent the radii of the inner and outer boundaries of the gravastar, respectively.
Consequently, the thickness of the intermediate shell region can be expressed as $\delta \equiv r_2 - r_1$,
where $\delta$ could be much smaller than the mass parameter of the exterior Schwarzschild spacetime and the length parameter of the interior de Sitter spacetime.

The concept of gravastars as an alternative to black holes has been extensively explored in the literature~\cite{Visser:2003ge, Cattoen:2005he, Carter:2005pi,
Bilic:2005sn, Lobo:2005uf, DeBenedictis:2005vp, Lobo:2006xt, Horvat:2007qa, Chirenti:2007mk, Rocha:2008hi, Horvat:2008ch,
Usmani:2008ce, Turimov:2009mu, Nandi:2008ij, Harko:2009gc, Usmani:2010ac, Rahaman:2011we, Rahaman:2012wc, Bhar:2014vra, Rahaman:2012xx,
Ghosh:2015ohi, Ghosh:2017flc, Ghosh:2019nsi, Ghosh:2019upa, Chan:2010fs}.
Mazur and Mottola proposed a gravastar model with a three-region structure and examined its thermodynamic stability through entropy maximization methods.
Meanwhile, Visser~\cite{Visser:2003ge} illustrated the dynamic stability of this system based on the three-region model,
It is currently well-established that the theoretical viability of the gravastar model, whether in varying dimensions or with or without charge, holds
significant importance within the realm of astrophysics.
Examining the Vaidya exterior spacetime, Chan et al.~\cite{Chan:2011wi} investigated the dynamic models of prototype radiating gravastars.
Their research unveiled that the final out could materialize as a range of entities, such as a black hole, an unsteady gravastar, a stable gravastar,
or a ``bounded excursion'' gravastar.
The particular result depends on variables like the evolving shell mass, the cosmological constant, and the initial placement of the moving shell.
Such inquiries are primarily investigated within the context of Albert Einstein's general relativity (GR).
Although it is widely acknowledged that GR remains one of the most encouraging theories for uncovering many of nature's enigmas,
it has become clear that an alteration of GR is necessary to overcome its shortcomings in both the theoretical and observational realms.
The confirmation of the universe's acceleration, along with the existence of dark matter and dark energy, as indicated by specific observational discoveries,
presents a theoretical challenge to GR \cite{SupernovaSearchTeam:1998fmf, SupernovaCosmologyProject:1998vns,
Boomerang:2000efg, Hanany:2000qf, Peebles:2002gy, Padmanabhan:2002ji, Clifton:2011jh, Riess:2006fw, SDSS:2003eyi, Amanullah:2010vv, WMAP:2010qai}.
In order to account for the current acceleration of the universe, several successive alternative gravity theories have emerged.
These theories include $f(T)$ gravity, $f(R)$ gravity, $f(R, \mathcal{T})$ gravity, and $f(T, \mathcal{T})$ gravity,
where $\mathcal{T}$ is the trace of the energy-momentum tensor, each of which incorporates modifications either
in the geometric aspect or in the matter-energy component of the Einstein field equations.
In the majority of such gravitational theories, various functions of $R$, the Ricci scalar, either alone or in conjunction with other scalar quantities, are integrated
into the gravitational Lagrangian within the corresponding action.
Numerous studies have been conducted on compact stars and gravastars within the context of amended theories,
as documented in \cite{Das:2015gwa, Das:2016mxq, Das:2017rhi, Biswas:2018inc, Ghosh:2020rau, Das:2020lqy, Sengupta:2020lhw, Banerjee:2020zjq}.
However, in the majority of such gravitational theories, the energy-momentum tensor $T_{\mu \nu}$ does not exhibit general conservation properties, in contrast to the scenario in GR.
This intriguing facet of modified gravity serves as our motivation to conduct a comprehensive examination of gravastars within the frame
of the GR theory with two scalars, formulated in \cite{Nojiri:2020blr}.
In the model, the scalar fields often become ghosts, which have kinetic energies unbounded below as the classical theory and generate negative norm states in the
quantum theory.
In order to avoid the ghosts, we impose constraints to make the scalar fields become non-dynamical, that is, so that the scalar fields do not propagate.
In this paper, we often use the word ``non-dynamical'' in meaning that any perturbation of the fields from the background does not propagate,
which is true even if we consider the time-dependent background. 
The constraints are similar to the mimetic one proposed in \cite{Chamseddine:2014vna}, where non-dynamical dark matter is generated.
The constraints make the solution in the present model stable.
Due to this stability, we can realize the stable gravastar with large surface redshift which was prohibited by the stability
in the previous works \cite{Buchdahl:1959zz, Straumann:1984xf, Boehmer:2006ye, Ivanov:2002xf, Barraco:2002ds, Boehmer:2006ye}.

The structure of this current study unfolds as follows:
In Section~\ref{S2}, we elucidate the fundamental mathematical framework of the Einstein GR theory with two scalar fields and provide the specific expressions for the field equations.
We impose the constraints for the scalar fields so that the scalar fields become non-dynamical.
We show that any spherically symmetric solution in the present model is always stable under any fluctuation because the two scalar fields become non-dynamical.
In Section~\ref{S3}, we derive the model, which generates solutions satisfying the boundary conditions between the shell region and the interior and exterior regions of the gravastar.
In Section~\ref{S4}, we discuss the physics of the gravastar by driving the proper length and the surface redshift of the shell.
Here we show that the model in this paper realizes the stable gravastar with large surface redshift prohibited due to the instability in the previous works.
In Section~\ref{S5}, we wrap up our study with a summary and conclusion.

\section{Einstein gravity coupled with two scalar fields}\label{S2}

Einstein's GR with two scalar fields $\phi$ and $\chi$ is described by the action as follows~\cite{Nojiri:2020blr},
\begin{align}
\label{I8}
S_{\mathrm{GR} \phi\chi} = \int d^4 x \sqrt{-g} & \left[ \frac{R}{2\kappa^2}
 - \frac{1}{2} \, A (\phi,\chi) \partial_\mu \phi \partial^\mu \phi
 - B (\phi,\chi) \, \partial_\mu \phi \partial^\mu \chi \right. \nn 
& \left. \quad - \frac{1}{2} \, C (\phi,\chi) \partial_\mu \chi \partial^\mu \chi - V (\phi,\chi)\right] \, .
\end{align}
In this context, $g$ represents the determinant of the metric tensor $g_{\mu\nu}$, $R$ denotes the Ricci scalar, and $V(\phi, \chi)$ represents the potential of the scalar doublet.
The functional forms of the coefficient functions $A$, $B$, and $C$ are contingent upon the properties of the scalars.
Upon varying the action~(\ref{I8}) with respect to the metric $g_{\mu\nu}$, we derive the ensuing Einstein equation,
\begin{align}
\label{I9}
\frac{1}{\kappa^2} \left( R_{\mu\nu} - \frac{1}{2} g_{\mu\nu} R \right) = &\, A (\phi,\chi) \partial_\mu \phi \partial_\nu \phi
+ B (\phi,\chi) \left( \partial_\mu \phi \partial_\nu \chi + \partial_\nu \phi \partial_\mu \chi \right)
+ C (\phi,\chi) \partial_\mu \chi \partial_\nu \chi\nn
&\, -g_{\mu\nu} \left[
\frac{1}{2}\, A (\phi,\chi) \partial_\rho \phi \partial^\rho \phi
+ B (\phi,\chi) \partial_\rho \phi \partial^\rho \chi
 + \frac{1}{2} \, C (\phi,\chi) \partial_\rho \chi \partial^\rho \chi + V (\phi,\chi)\right] \, .
\end{align}
Through the variation of action~(\ref{I8}) concerning the scalar fields $\phi$ and $\chi$, we acquire the following expressions,
\begin{align}
\label{I10}
0 =&\, \frac{A_\phi}{2}\, \partial_\mu \phi \partial^\mu \phi
+ A \nabla^\mu \partial_\mu \phi + A_\chi \partial_\mu \phi \partial^\mu \chi
+ \left( B_\chi - \frac{1}{2} \, C_\phi \right)\partial_\mu \chi \partial^\mu \chi + B \nabla^\mu \partial_\mu \chi - V_\phi \, ,\\
\label{I10b}
0 =&\, \left( - \frac{1}{2} \, A_\chi + B_\phi \right)
\partial_\mu \phi \partial^\mu \phi + B \nabla^\mu \partial_\mu \phi + \frac{1}{2} \, C_\chi \partial_\mu \chi \partial^\mu \chi
+ C \nabla^\mu \partial_\mu \chi + C_\phi \partial_\mu \phi \partial^\mu \chi - V_\chi\, .
\end{align}
Here, let us define $A_\phi$ as $\partial A(\phi,\chi)/\partial \phi$, and similarly for other derivatives.
It is worth noting that Eqs.~(\ref{I10}) and (\ref{I10b}) can be derived from the Bianchi identity in conjunction with Eq.~(\ref{I9}).

In the following, we identify
\begin{align}
\label{TSBH1}
\phi=t\, , \quad \chi=r\, .
\end{align}
As explained in the reference \cite{Nojiri:2020blr}, making the assumption (\ref{TSBH1}) does not result in any loss of generality.
In the case of a spacetime with a general spherically symmetric yet time-dependent solution, the scalar fields $\phi$ and $\chi$ exhibit dependencies
on both the time coordinate, denoted as $t$ and the radial coordinate, denoted as $r$.
In the context of a given solution, the specific dependencies of $\phi$ and $\chi$ on both the time variable $t$
and the radial variable $r$ are determined as functions: $\phi = \phi(t, r)$ and $\chi = \chi(t, r)$.
We may redefine the scalar fields to replace $t$ and $r$ with new scalar fields, $\tilde\phi$ and $\tilde\chi$,
$\phi\left( \tilde\phi, \tilde\chi \right) \equiv \phi\left( t=\tilde\phi, r=\tilde\chi \right)$ and
$\chi\left( \tilde\phi, \tilde\chi \right) \equiv \chi\left( t=\tilde\phi, r=\tilde\chi \right)$.
Subsequently, we can associate the new scalar fields with the time and radial coordinates in (\ref{TSBH1}).
The transformation of variables from $\left(\phi,\chi\right)$ to $\left(\tilde\phi,\tilde\chi\right)$ can be absorbed
into the redefinitions of $A$, $B$, $C$, and $V$ within the action (\ref{I8}).
This demonstrates that making the assumption (\ref{TSBH1}) does not lead to any loss of generality.
Furthermore, as we will observe, $\phi$ is identified with $t$ even for a static spacetime.

If we assume the static and spherically symmetric spacetime, whose metric is given by
\begin{align}
\label{GBiv}
ds^2 = - \e^{2\nu (r)} dt^2 + \e^{2\mu (r)} dr^2 + r^2 \left( d\theta^2 + \sin^2\theta \, d\phi^2 \right)\, ,
\end{align}
the $\left( t,t \right)$, $\left(r,r\right)$, $\left(i,j\right)$, and $\left(t,r \right)$ components of the Einstein equation under the identification (\ref{TSBH1}) have the following forms,
\begin{align}
\label{TSBH2}
\frac{\e^{2\left(\nu - \mu\right)}}{\kappa^2} \left( \frac{2\mu'} r + \frac{\e^{2\mu} - 1}{r^2} \right)
=&\, - \e^{2\nu} \left( - \frac{A}{2} \e^{-2\nu} - \frac{C}{2} \e^{-2\mu} - V \right) \, ,\\
\label{TSBH3}
\frac{1}{\kappa^2} \left( \frac{2\nu'} r - \frac{\e^{2\mu} - 1}{r^2} \right) =&\, \e^{2\mu} \left( \frac{A}{2} \, \e^{-2\nu}
+ \frac{C}{2} \, \e^{-2\mu} - V \right) \, ,\\
\label{TSBH4}
\frac{1}{\kappa^2} \left[ \e^{-2\mu}\left( r \left(\nu' - \mu' \right) + r^2 \nu'' + r^2 \left( \nu' - \mu' \right) \nu' \right) \right]
=&\, r^2 \left( \frac{A}{2} \e^{-2\nu} - \frac{C}{2} \e^{-2\mu} - V \right) \, ,\\
\label{TSBH5}
0=&\, B \, ,
\end{align}
which can be solved with respect to $A$, $B$, $C$, and $V$, as follows,
\begin{align}
\label{TSBH6st}
A(r)=& \,\frac{\e^{2\left(\nu - \mu\right)}}{\kappa^2} \left[ \frac{\e^{2\mu} - 1}{r^2}
+ \frac{\nu' + \mu'} r + \nu'' + \left( \nu' - \mu' \right) \nu' \right] \, , \\
\label{TSBH7st}
B(r)=& \, 0 \, , \\
\label{TSBH8st}
C(r)=& \, \frac{1}{\kappa^2} \left[ - \frac{\e^{2\mu} - 1}{r^2} + \frac{\nu' + \mu'} r - \nu'' - \left( \nu' - \mu' \right) \nu' \right] \, , \\
\label{TSBH9st}
V(r)=& \, \frac{\e^{-2\mu}}{2\kappa^2} \left[ \frac{2\left( \mu' - \nu'\right) } r + \frac{2 \left(\e^{2\mu} - 1\right) }{r^2} \right] \, .
\end{align}
This tells that if we consider the model where the radial coordinate $r$ in $A$, $B$, $C$, and $V$ of Eqs.~(\ref{TSBH6st}), (\ref{TSBH7st}), (\ref{TSBH8st}), and (\ref{TSBH9st}),
is replaced with $\chi$, that is, $A(r=\chi)$, $B(r=\chi)$, $C(r=\chi)$, and $V(r=\chi)$, the spacetime given by Eq.~(\ref{GBiv}) becomes a solution of the model.

We should note the product of $A$ and $C$ is always negative, which tells us that either $\phi$ or $\chi$ is a ghost.
The ghost mode has negative kinetic energy classically and generates negative norm states as a quantum theory.
Therefore the existence of the ghost mode tells that the model is physically inconsistent.
We now eliminate ghosts by imposing constraints on $\phi$ and $\chi$, which is similar to the mimetic constraint in \cite{Chamseddine:2013kea}, and make the ghost modes non-dynamical.
For the purpose, by using the Lagrange multiplier fields $\lambda_\phi$ and $\lambda_\chi$, we add the following terms to the action $S_{\mathrm{GR} \phi\chi}$ in (\ref{I8})
as $S_{\mathrm{GR} \phi\chi} \to S_{\mathrm{GR} \phi\chi} + S_\lambda$,
\begin{align}
\label{lambda1}
S_\lambda = \int d^4 x \sqrt{-g} \left[ \lambda_\phi \left( \e^{-2\nu(r=\chi)} \partial_\mu \phi \partial^\mu \phi + 1 \right)
+ \lambda_\chi \left( \e^{-2\mu(r=\chi)} \partial_\mu \chi \partial^\mu \chi - 1 \right) \right] \, .
\end{align}
The variations of $S_\lambda$ with respect to $\lambda_\phi$ and $\lambda_\chi$ give the constraints\footnote{
The constraint given by Eq.~(\ref{lambda2}), is similar to the constraint of mimetic theory; see,
for example, \cite{Chamseddine:2014vna, Nojiri:2022cah,Myrzakulov:2015kda, Nashed:2021hgn, Nashed:2023jdf, Nashed:2023fzp,
Myrzakulov:2015qaa,Vagnozzi:2017ilo,Nashed:2021ctg, Casalino:2018tcd,Casalino:2018wnc,Sebastiani:2016ras,Nashed:2021pkc}.}
\begin{align}
\label{lambda2}
0 = \e^{-2\nu(r=\chi)} \partial_\mu \phi \partial^\mu \phi + 1 \, , \quad
0 = \e^{-2\mu(r=\chi)} \partial_\mu \chi \partial^\mu \chi - 1 \, ,
\end{align}
whose solutions are consistently given by (\ref{TSBH1}).
We should note that even in the model modified by adding $S_\lambda$, $\lambda_\phi=\lambda_\chi=0$ is always a solution.
Therefore the spacetime given by Eq.~(\ref{GBiv}) is a solution even in the modified theory.
The constraints in (\ref{lambda2}) make the scalar field $\phi$ and $\chi$ non-dynamical, that is, the fluctuation of $\phi$ and $\chi$ from the background (\ref{TSBH1})
do not propagate.
In fact, by considering the perturbation from (\ref{TSBH1}),
\begin{align}
\label{pert1}
\phi=t + \delta \phi \, , \quad \chi=r + \delta \chi\, ,
\end{align}
we obtain,
\begin{align}
\label{pert2}
\partial_t \delta \phi = \partial_r \delta \chi = 0\, .
\end{align}
Therefore if we impose the initial condition $\delta\phi=0$, we find $\delta\phi=0$ in the whole spacetime.
On the other hand, if we impose the boundary condition $\delta\chi\to 0$ when $r\to 0$, we find $\delta\chi=0$ in the whole spacetime.
These tell that $\phi$ and $\chi$ are non-dynamical.
Therefore the solution given by the spherically symmetric spacetime in (\ref{GBiv}) is always stable under the perturbation of the scalar fields.
Only the propagating mode or possible fluctuation in this model is given by the gravitational wave.
Because the metric minimally couples with scalar fields, the propagation of the gravitational wave does not change from the standard one,
and therefore the fluctuation does not cause any instability, and the solution in (\ref{GBiv}) is always stable under any fluctuation.
This situation is similar to the original mimetic gravity theory in \cite{Chamseddine:2013kea}, where effective dark matter appeared.
The effective dark matter is not dynamical and there does not appear any fluctuation.
For example, the effective dark matter never collapses due to gravity.
Therefore the formulation by using the constraints in (\ref{lambda2}) could be regarded as a natural extension of the original mimetic gravity.

We often consider the spacetime where $\mu = - \nu$.
For the spacetime, Eqs.~(\ref{TSBH6st}), (\ref{TSBH7st}), (\ref{TSBH8st}), and (\ref{TSBH9st}) are reduced as follows,
\begin{align}
\label{TSBHst2}
A=&\, \frac{\e^{2\nu}}{\kappa^2} \left[ - \frac{\e^{2\nu} - 1}{r^2} + \frac{1}{2} \left(\e^{2\nu}\right)''\right] \, , \quad
B= 0 \, , \nonumber \\
C=&\, \frac{\e^{-2\nu}}{\kappa^2} \left[ \frac{\e^{2\nu} - 1}{r^2} - \frac{1}{2} \left(\e^{2\nu}\right)'' \right] \, , \quad
V= \frac{1}{\kappa^2} \left[ - \frac{\left( \e^{2\nu} \right)'} r - \frac{\e^{2\nu} - 1}{r^2} \right] \, .
\end{align}
We use the above expressions in the following.

\section{Realization of Gravastar geometry}\label{S3}

We now construct the model which realizes gravastar spacetime.
In the gravastar geometry, there is an exterior region which is the Schwarzschild spacetime where
\begin{align}
\label{Sch1}
\e^{2\nu}=\e^{-2\mu} = 1 - \frac{2M} r \, ,
\end{align}
and the interior region is described by the de Sitter spacetime,
\begin{align}
\label{dS1}
\e^{2\nu}=\e^{-2\mu} = 1 - \lambda^2 r^2 \, .
\end{align}
Here $M$ corresponds to the mass measured by the observer at infinity.
On the other hand, $\lambda^2$ corresponds to the cosmological constant.
The exterior region and the interior region are connected by the shell region.
We now consider the two cases.
In one case, $\e^{2\nu}=\e^{-2\mu}$ in the shell region is given by the polynomial of $r$.
In another case, we try to connect the two regions by using smooth functions, which describe the asymptotically Schwarzschild spacetime
when $r$ is large but describe the asymptotically de Sitter spacetime when $r$ is small.

\subsection{Shell model}

We consider the shell region in $r_1<r<r_2$, with a boundary with the interior region at $r=r_1$ and the boundary with the exterior region at $r=r_2$.
We assume that $\e^{2\nu}=\e^{-2\mu}$ in the shell region is given by the following polynomial,
\begin{align}
\label{fr}
\e^{2\nu}=\e^{-2\mu}= f(r) = a_0 + a_1 \left( r - r_1 \right) + a_2 \left( r - r_1 \right)^2 + a_3 \left( r - r_1 \right)^3 + a_4 \left( r - r_1 \right)^4 + a_5 \left( r - r_1 \right)^5 \, .
\end{align}
By imposing the continuities of $\e^{2\nu}$, $\left(\e^{2\nu}\right)'$, and $\left(\e^{2\nu}\right)''$ at the two boundaries, we obtain,
\begin{align}
\label{as1}
a_0=&\, 1 - \lambda^2 {r_1}^2 >0 \, , \quad a_1= - 2 \lambda^2 r_1 <0\, , \quad a_2 = - \lambda^2 <0 \, , \nonumber \\
a_3 =&\, - \frac{1}{{r_2}^3\left(r_2 -r_1 \right)^3} \left\{ 2 \left( 15{r_2}^2 - 6 r_2 r_1 + {r_1}^2\right)M - {r_2}^3 \left( 3{r_2}^2 + 6 r_2 r_1 + {r_1}^2 \right) \lambda^2 \right\} \, , \nonumber \\
a_4 =&\, \frac{1}{{r_2}^3 \left( r_2 - r_1 \right)^4} \left\{ 2 \left( 24 {r_2}^2 - 11 r_2 r_1 + 2{r_1}^2 \right) M - {r_2}^3 \left( 3{r_2}^3 + 10 r_2 r_1 + 2 {r_1}^2 \right) \lambda^2 \right\} \, , \nonumber \\
a_5 =&\, - \frac{1}{{r_2}^3 \left( r_2 - r_1 \right)^5} \left\{ 2 \left( 10 {r_2}^2 - 5 r_2 r_1 + {r_1}^2 \right) M - {r_2}^3\left( {r_2}^2 + 4 r_2 r_1 + {r_1}^2 \right) \lambda^2 \right\} \, .
\end{align}
Because $\frac{{r_2}^2 + 4 r_2 r_1 + {r_1}^2}{10 {r_2}^2 - 5 r_2 r_1 + {r_1}^2}>0$, we may choose $M$ by
\begin{align}
\label{Mchosen}
M= \frac{{r_2}^3 \left( {r_2}^2 + 4 r_2 r_1 + {r_1}^2 \right) \lambda^2}{2\left( 10 {r_2}^2 - 5 r_2 r_1 + {r_1}^2 \right)}\, ,
\end{align}
so that $a_5=0$.
In this choice, we find
\begin{align}
\label{as2}
a_0=&\, 1 - \lambda^2 {r_1}^2 >0 \, , \quad a_1= - 2 \lambda^2 r_1 <0\, , \quad a_2 = - \lambda^2 <0 \, , \nonumber \\
a_3 =&\, \frac{\lambda^2 r_2 \left( 15 {r_2}^2 + 6 r_2 r_1 -3 {r_1}^2 \right)}{\left(r_2 -r_1 \right)^2 \left( 10 {r_2}^2 - 5 r_2 r_1 + {r_1}^2 \right)} >0 \, , \nonumber \\
a_4 =&\, - \frac{3\lambda^2 r_2 \left( 2 {r_2}^2 + 2 r_2 r_1 - {r_1}^2 \right)}{\left( r_2 - r_1 \right)^3\left( 10 {r_2}^2 - 5 r_2 r_1 + {r_1}^2 \right)} < 0 \, . \nonumber \\
\end{align}
We also find
\begin{align}
\label{derfs}
f(r_1)=&\, a_0 >0 \, , \quad f'(r_1) = a_1 <0\, , \quad f''(r_1) = 2 a_2 <0 \, , \quad f'''(r_1)= 6a_3 >0 \, . \quad
f''''(r) = 24 a_4 <0\, , \nonumber \\
f(r_2)=&\, 1 - \frac{2M}{r_2}>0\, , \quad f'(r_2) =\frac{2M}{{r_2}^2} >0 \, , \quad f''(r_2) = - \frac{4M}{{r_2}^3} <0\, , \nonumber \\
f'''(r_2)=&\, 6a_3 + 24 a_4 \left(r_2 -r_1 \right)
= - \frac{54\lambda^2 r_2 \left( {r_2}^2 + 2 r_2 r_1 - {r_1}^2 \right)}{\left(r_2 -r_1 \right)^2 \left( 10 {r_2}^2 - 5 r_2 r_1 + {r_1}^2 \right)} <0 \, .\nonumber \\
\end{align}
This tells the function $f(r)$ has two maxima when $r<r_1$ and $r>r_2$ and one minimum when $r_1<r<r_2$.
The values of $r$ corresponding to the extrema are given by solving the equations
\begin{align}
\label{extrema}
0=f'(r) = a_1 + 2 a_2 \left( r - r_1 \right) + 3 a_3 \left( r - r_1 \right)^2 + 4 a_4 \left( r - r_1 \right)^3 \, .
\end{align}
Because all $a_i$'s $\left(i=1,\cdots,4\right)$ are proportional to $\lambda^2$, the solution of (\ref{extrema}) does not depend on $\lambda$.
Therefore if we choose $\lambda$ to be sufficiently small, that is, $\lambda r_1$, $\lambda r_2 \ll 1$, only $a_0$ dominates $a_0\sim 1$ in the expression of $f(r)$ in (\ref{fr})
and therefore $f(r)$ can be consistently positive.
If $f(r)$ could become negative, the horizon would appear and therefore the solution would not describe the gravastar but a black hole, which could be regular at the origin.
We should also note $f(r)<1$.
On the other hand, even if $\lambda$ is not so small, that is, $\lambda r_1$, $\lambda r_2 \sim 1$, when the thickness of the shell is small enough, $\lambda \left( r_2 - r_1 \right) \ll 1$,
$f(r)$ could be positive.

\

\begin{figure}
\centering
\subfigure[~The behavior of the polynomial function $f(r)$ given by Eq.~(\ref{fr}) ]{\label{fig:f}\includegraphics[scale=0.3]{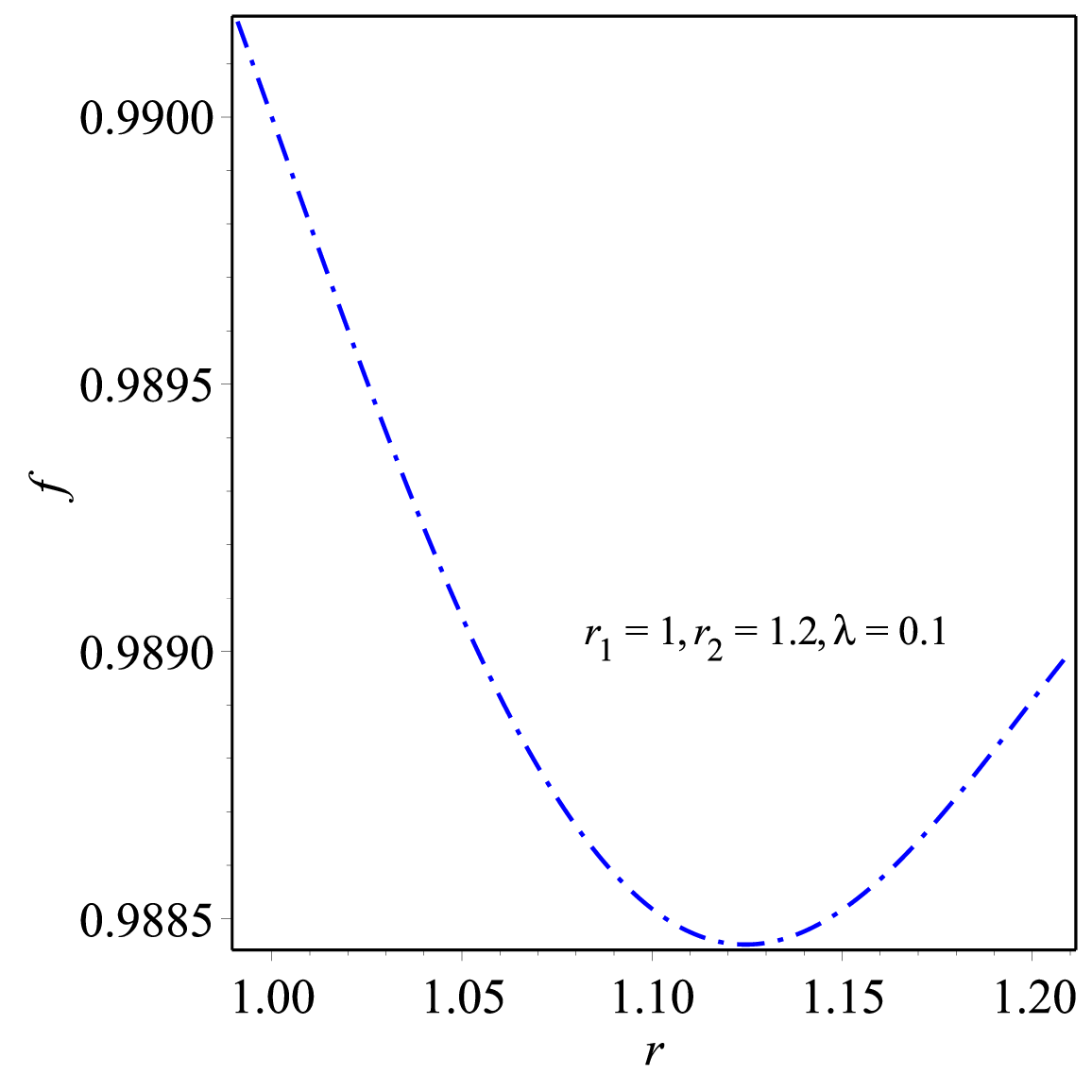}}
\subfigure[~The behavior of the scalar field $A(r)$ given by Eqs.~(\ref{scalar1}) ]{\label{fig:A}\includegraphics[scale=0.3]{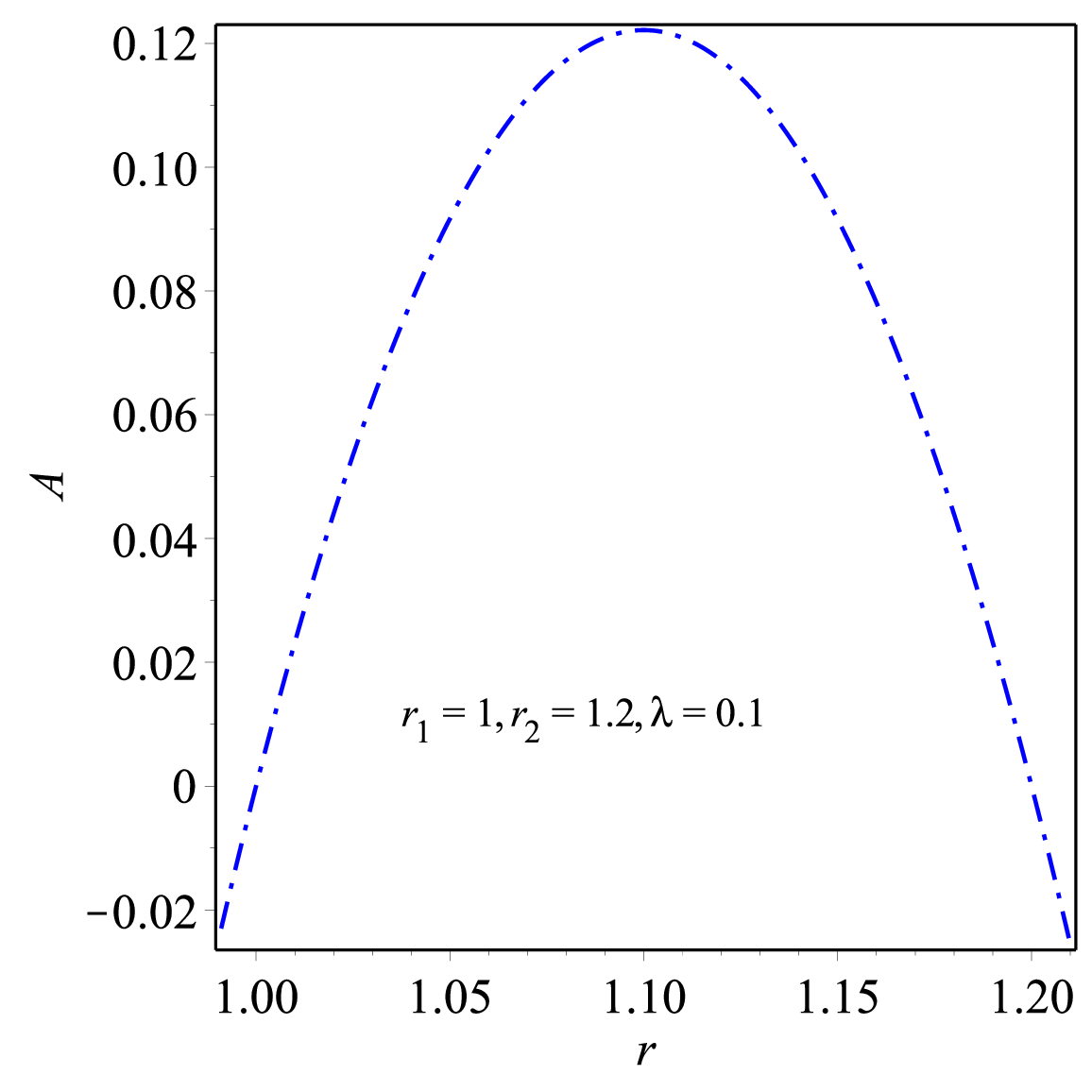}}
\subfigure[~The behavior of the scalar field $C(r)$ given by Eqs.~(\ref{scalar1}) ]{\label{fig:C}\includegraphics[scale=0.3]{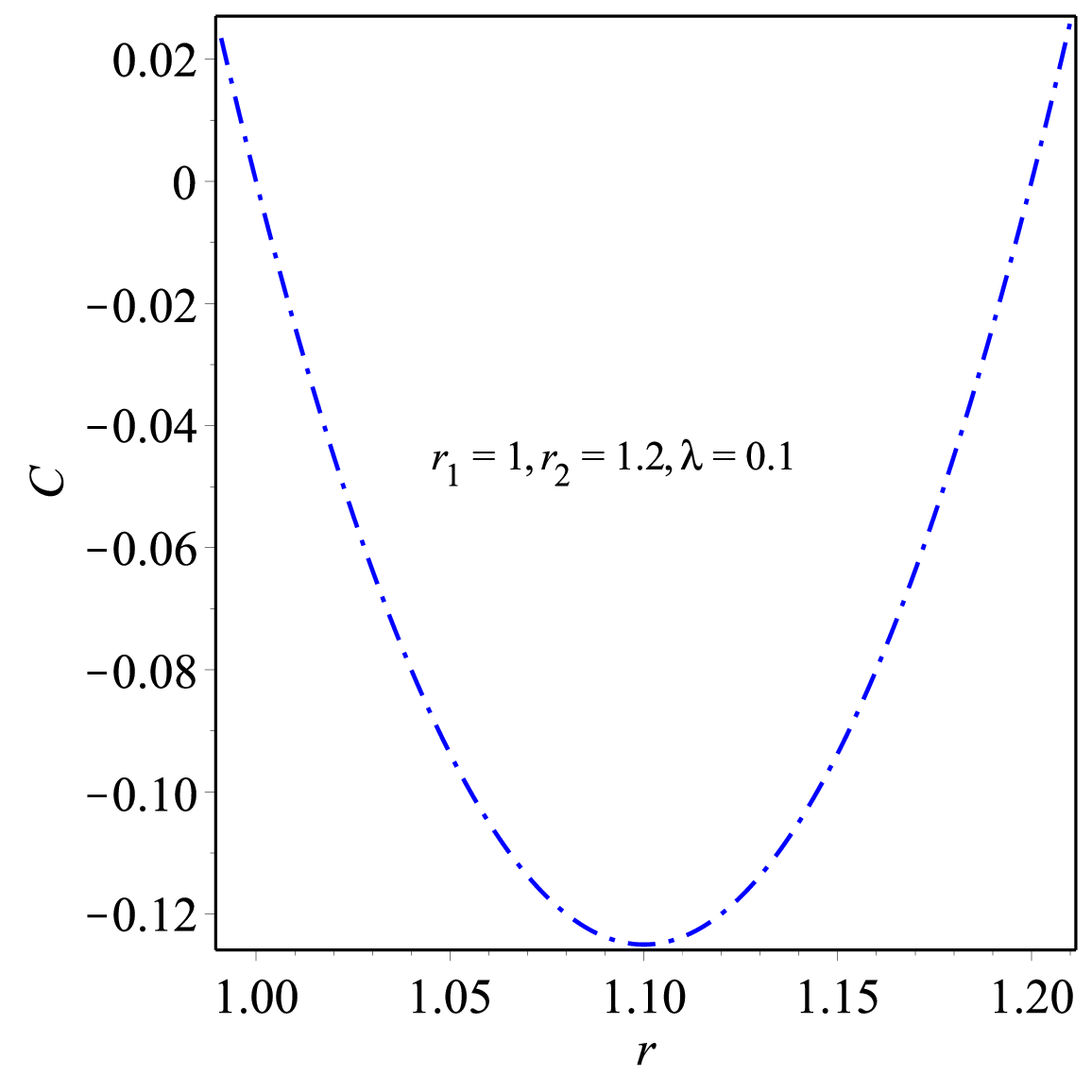}}
\subfigure[~The behavior of the potential field $V(r)$ given by Eqs.~(\ref{scalar1}) ]{\label{fig:V}\includegraphics[scale=0.3]{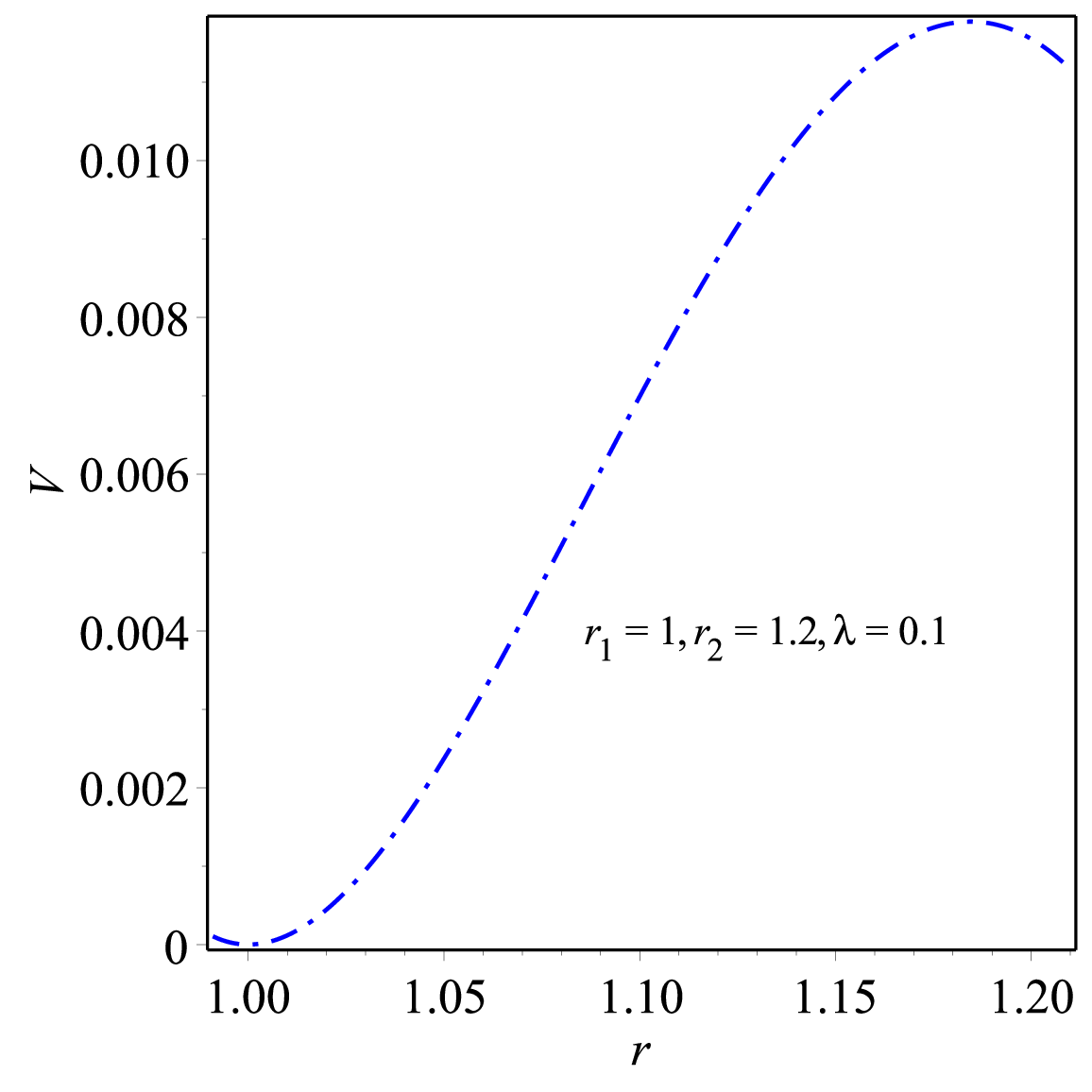}}
\caption[figtopcap]{\small{Plots of Figures in the case of small $\lambda$ (\ref{Poly1}). \subref{fig:f} the polynomial function given by Eq.~(\ref{fr})
after using Eq.~(\ref{as2}); Figures~\subref{fig:A} \subref{fig:C}, and \subref{fig:V} show the behavior of the scalars $A(r)$, $C(r)$, and the potential $V(r)$ given by Eqs.~(\ref{scalar1}).}}
\label{Fig:1}
\end{figure}

\begin{figure}
\centering
\subfigure[~The behavior of the polynomial function $f(r)$ given by Eq.~(\ref{fr}) ]{\label{fig:f2}\includegraphics[scale=0.3]{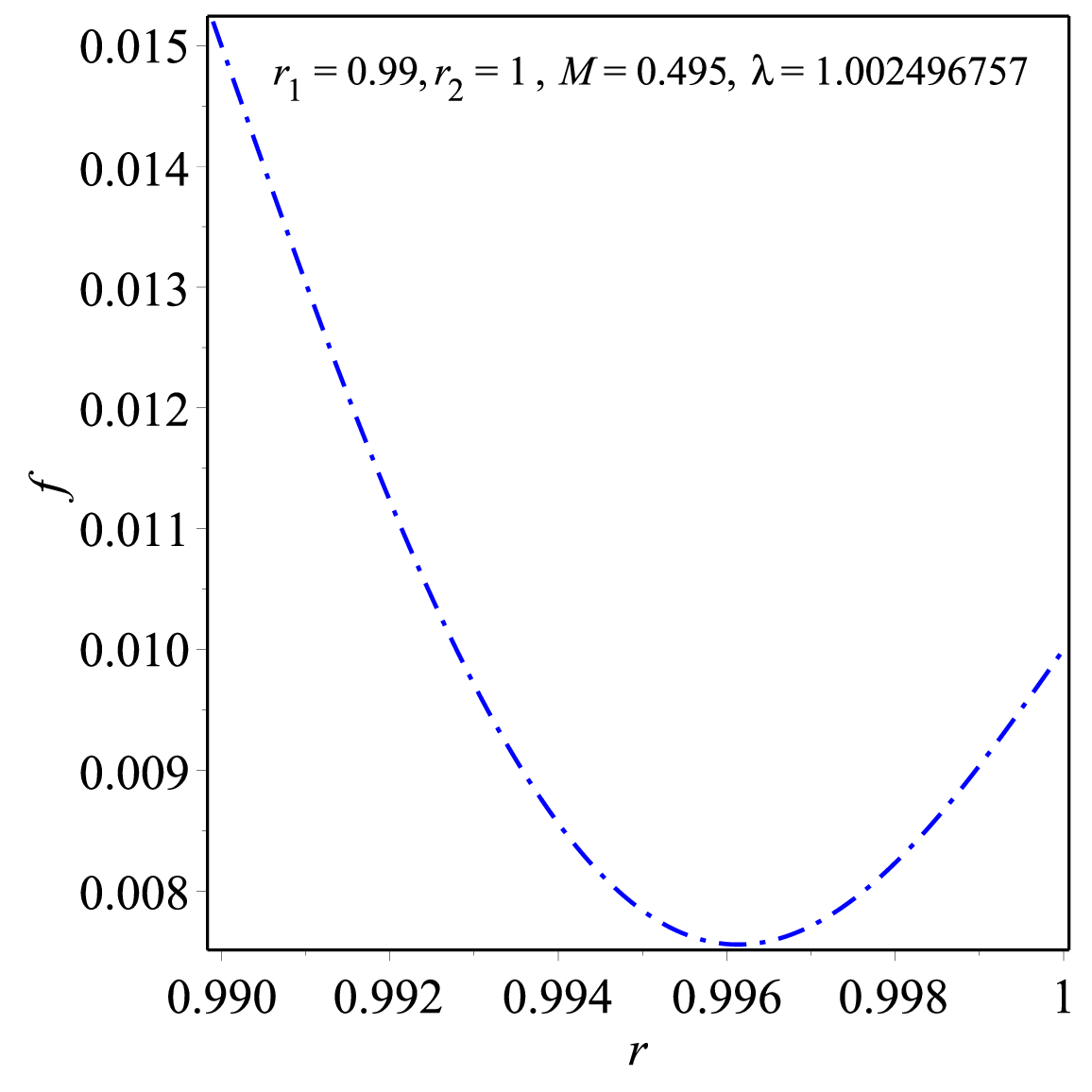}}
\subfigure[~The behavior of the scalar field $A(r)$ given by Eqs.~(\ref{scalar1}) ]{\label{fig:A2}\includegraphics[scale=0.3]{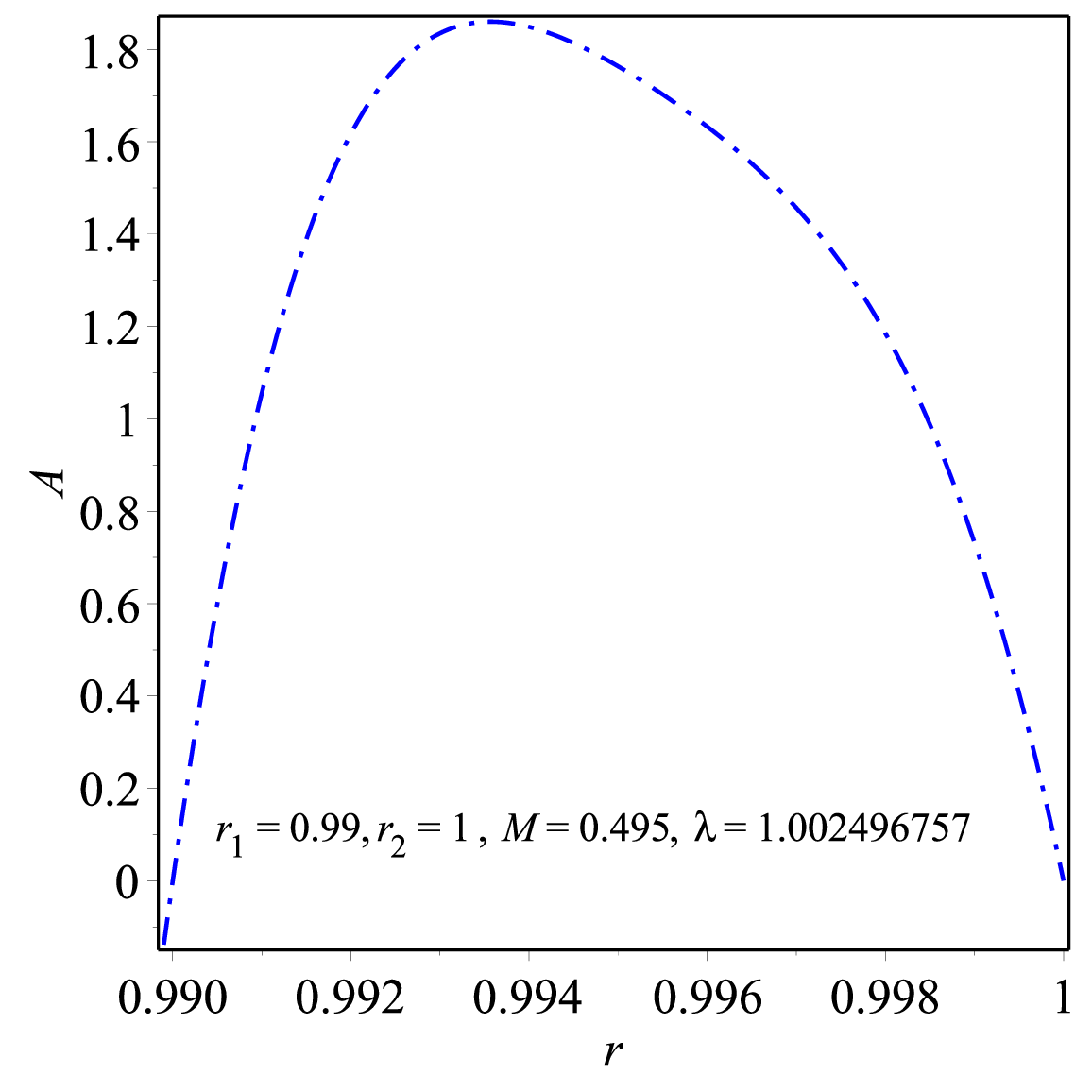}}
\subfigure[~The behavior of the scalar field $C(r)$ given by Eqs.~(\ref{scalar1}) ]{\label{fig:C2}\includegraphics[scale=0.3]{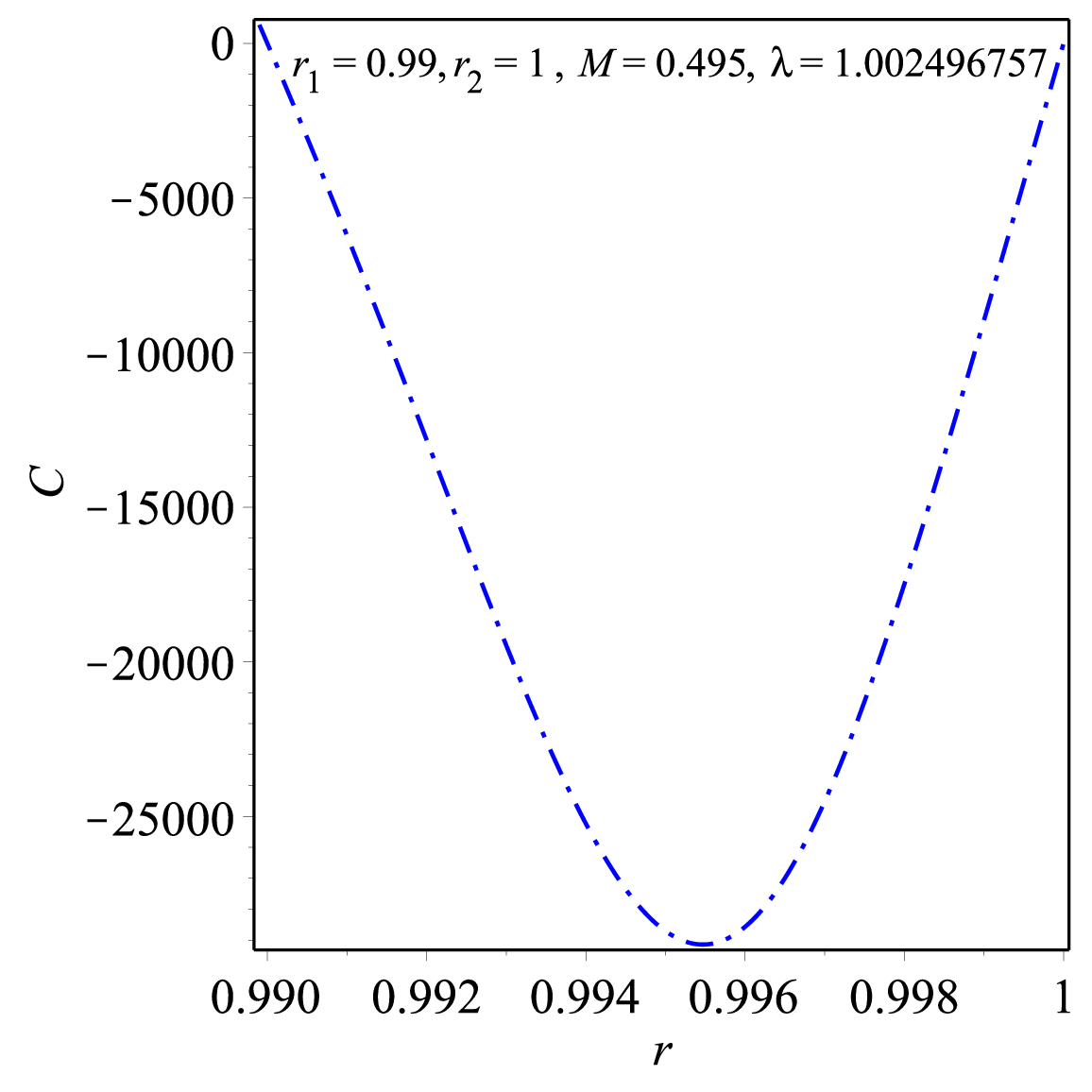}}
\subfigure[~The behavior of the potential field $V(r)$ given by Eqs.~(\ref{scalar1}) ]{\label{fig:V2}\includegraphics[scale=0.3]{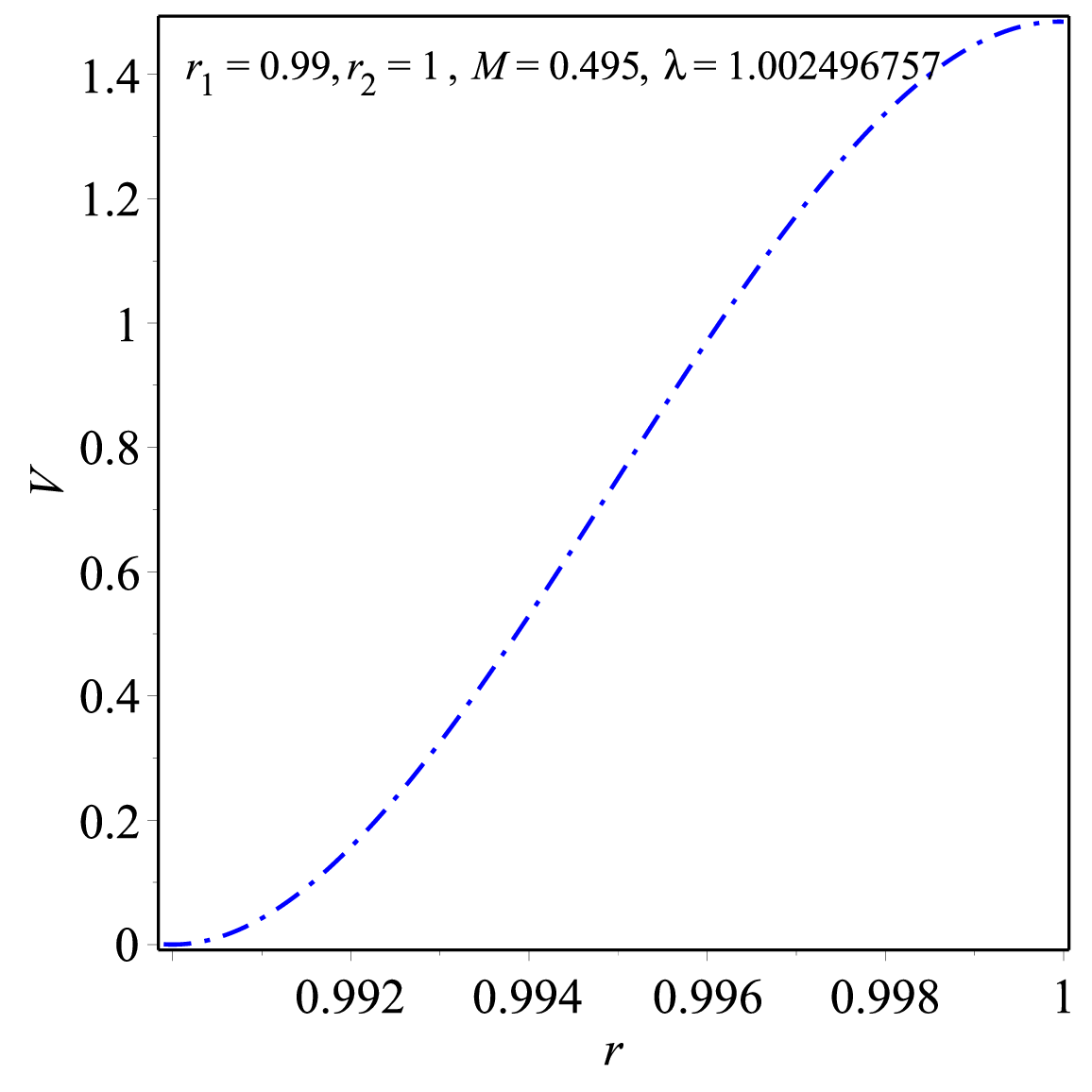}}
\caption[figtopcap]{\small{Plots of Figures in the case of thin shell (\ref{Poly1}). \subref{fig:f} the polynomial function given by Eq.~(\ref{fr})
after using Eq.~(\ref{as2}); Figures~\subref{fig:A} \subref{fig:C}, and \subref{fig:V} show the behavior of the scalars $A(r)$, $C(r)$, and the potential $V(r)$ given by Eqs.~(\ref{scalar1}).}}
\label{Fig:1B}
\end{figure}

We numerically investigate typical two cases, where we choose,
\begin{itemize}
\item Sufficiently small $\lambda$ case,
\begin{align}
\label{Poly1}
r_1=1\, , \quad r_2=1.2\, , \quad \lambda=0.1\, , \quad \mbox{which gives} \quad M=0.000665 \cdots\, .
\end{align}
The dimension of the mass or length is fixed by choosing $r_1=1$.
When we include the dimension, $\lambda r_{1,2}$ and $\lambda M$ are dimensionless.
\item Sufficiently thin shell case, where we choose,
\begin{align}
\label{Poly2}
r_1=0.99\, , \quad r_2=1\, , \quad M=0.495\, , \quad \mbox{which gives} \quad \lambda = 1.002496757\cdots \, .
\end{align}
Here the dimension of the mass or length is fixed by choosing $r_2=1$.
\end{itemize}
The behaviors of $f(r)$ in Eq.~(\ref{fr}) are shown in Figure~\ref{Fig:1}~\subref{fig:f} for the case of (\ref{Poly1}) and Figure~\ref{Fig:1B}~\subref{fig:f2}
for the case of (\ref{Poly2}).
In the case of (\ref{Poly1}) , we find $M\ll r_2$ but in the case of (\ref{Poly2}), $r_2\sim 2M$.
Note that $r_2$ corresponds to the radius of the surface in the gravastar.
Therefore in the case of (\ref{Poly2}), the radius of the gravastar is smaller than the radius of the photon sphere $\sim 6M$
and the photon sphere could be observed in the case of (\ref{Poly2}), which could be consistent with the observation
of the Event Horizon Telescope~\cite{EventHorizonTelescope:2019ths}.
As we will show later, in the case of (\ref{Poly2}), the surface redshift becomes very large because $r_2\sim 2M$
and the redshift lies inside the region prohibited by the stability in the previous works.

In the interior region $\left(r<r_1\right)$, we find
\begin{align}
\label{ABCVinterior}
A=B=C=0\, , \quad V=\frac{3\lambda^2}{\kappa^2}\, ,
\end{align}
which is consistent with the fact that the interior region is the de Sitter spacetime.
On the other hand, in the exterior region $\left(r>r_2\right)$, we obtain,
\begin{align}
\label{ABCVexterior}
A=B=C=V=0\, ,
\end{align}
which corresponds to the Schwarzschild spacetime which is a vacuum solution of the Einstein equation.
In the shell region $\left(r_1<r<r_2\right)$, $A$, $B$, $C$, and $V$ are given by substituting (\ref{fr}) with (\ref{as1}) or (\ref{as2}) into the expressions in (\ref{TSBHst2}),
\begin{align}\label{scalar1}
A=&\, 3 \lambda^2r_2 \left\{ 1 + \lambda^2 \left( - {r_1}^2 - 2 r_1 \left( r-r_1 \right) - \left( r-r_1 \right)^2
+ \frac{ 3 \left( 5 {r_2}^2 + 2 r_2 r_1 - {r_1}^2 \right) r_2 \left( r-r_1 \right)^3}{ \left( r_2-r_1 \right)^2 \left( 10 {r_2}^2 - 5 r_2\,r_1+{r_1}^2 \right) } \right. \right. \nonumber \\
&\, \left. \left. \qquad - \frac{ 3\left( 2 {r_2}^2 + 2 r_2 r_1 - {r_1}^2 \right) r_2 \left( r-r_1 \right)^4}{ \left( 10 {r_2}^2-5 r_2 r_1+{r_1}^2 \right) \left(r_2-r_1 \right)^3}
\right) \right\} \left\{ 5{r_2}^3{r_1}^3 - {r_2}^2{r_1}^4 -r_2 {r_1}^5 \right. \nonumber \\
&\, \qquad + \left({r_2}^2{r_1}^3 - 15 {r_2}^3{r_1}^2 + r_2\,{r_1}^4+{r_1}^5\right) r
+ \left(10 {r_2}^3 + 10 {r_2}^2 r_1 + 10 r_2 {r_1}^2 -6 {r_1}^3 \right)r^3 \nonumber \\
&\, \left. \qquad + \left( 5 {r_1}^2- 10 {r_2}^2 - 10 r_2 r_1 \right) r^4 \right\} \left( 10 {r_2}^2 - 5 r_2 r_1+{r_1}^2 \right)^{-1}
\left( r_2-r_1 \right)^{- 3} r^{-2}\,,\nonumber \\
C=&\, -3\lambda^2r_2\left\{ 5{r_2}^3{r_1}^3 - {r_2}^2{r_1}^4 - r_2 {r_1}^5
+ \left({r_1}^5- 15 {r_2}^3{r_1}^2 + {r_2}^2{r_1}^3+ r_2{r_1}^4 \right) r
\right. \nonumber \\
&\, \left. \qquad + \left(10 {r_2}^3 + 10{r_2}^2 r_1+ 10 r_2 {r_1}^2 - 6 {r_1}^3 \right) r^3 + \left( -10 {r_2}^2 + 5 {r_1}^2 - 10r_2r_1 \right) r^4 \right\} \nonumber \\
&\, \times \left\{ 1 + \lambda^2 \left( - {r_1}^2 - 2 r_1 \left( r-r_1 \right) - \left( r-r_1 \right)^2
+ \frac{ 3 \left( 5 {r_2}^2 + 2 r_2 r_1 - {r_1}^2 \right) r_2 \left( r-r_1 \right)^3}{ \left( r_2-r_1 \right)^2 \left( 10 {r_2}^2 - 5 r_2\,r_1+{r_1}^2 \right) } \right. \right. \nonumber \\
&\, \left. \left. \qquad - \frac{ 3\left( 2 {r_2}^2 + 2 r_2 r_1 - {r_1}^2 \right) r_2 \left( r-r_1 \right)^4}{ \left( 10 {r_2}^2-5 r_2 r_1+{r_1}^2 \right) \left(r_2-r_1 \right)^3}
\right) \right\}^{-1} \left( 10 {r_2}^2 - 5 r_2 r_1 + {r_1}^2 \right)^{-1} \left( r_2-r_1 \right)^{-3} r^{-2}\,,\nonumber \\
V=&\, 3 \lambda^2 r_2 \left\{ 10 {r_2}^3 {r_1}^3 - 2 {r_2}^2 {r_1}^4 - 2 r_2 {r_1}^5
+ \left( -15 {r_2}^3 {r_1}^2 + {r_2}^2 {r_1}^3 + r_2 {r_1}^4 + {r_1 }^5 \right) r \right. \nonumber \\
&\left. \qquad + \left( 5 {r_2}^3 + 5 {r_2}^2 r_1 + 5 r_2 {r_1}^2 - 3 {r_1}^3 \right) r^3
+ \left( - 4 {r_2}^2 - 4 r_2 r_1 + 2 {r_1}^2 \right) r^4 \right\} \nonumber \\
&\, \qquad \times \left\{ 2\left( 10\,{r_2}^2-5\,r_2 \,r_1+{r_1}^2 \right) \left( r_2-r_1 \right)^ {3} r^2\right\}^{-1}\,.
\end{align}
The above expressions are consistent with the interior and exterior solutions in (\ref{ABCVinterior}) and (\ref{ABCVexterior}) because
\begin{align}
\label{ABCVr1}
A\left(r_1\right) =&\, \frac{\e^{2\nu(r_1)}}{\kappa^2} \left[ - \frac{f\left(r_1\right) - 1}{{r_1}^2} + \frac{1}{2} f''\left(r_1\right)\right]
B\left(r_1\right) = 0 \, , \nonumber \\
C\left(r_1\right) =&\, \frac{\e^{-2\nu\left(r_1\right)}}{\kappa^2} \left[ \frac{f\left(r_1\right) - 1}{r^2} - \frac{1}{2} f''\left(r_1\right) \right]
= \frac{\e^{-2\nu\left(r_1\right)}}{\kappa^2} \left[ \frac{1 - \lambda^2{r_1}^2 - 1}{{r_1}^2} - \frac{1}{2} \left( - 2 \lambda^2 \right)\right] =0\, , \nonumber \\
V\left(r_1\right) =&\, \frac{1}{\kappa^2} \left[ - \frac{f'\left(r_1\right)}{r_1} - \frac{f\left(r_1\right) - 1}{{r_1}^2} \right]
= \frac{1}{\kappa^2} \left[ - \frac{- 2\lambda^2 r_1}{r_1} - \frac{1 - \lambda^2 {r_1}^2 - 1}{{r_1}^2} \right] = \frac{3\lambda^2}{\kappa^2}\, , \\
\label{ABCVr2}
A\left(r_2\right) =&\, \frac{\e^{2\nu(r_2)}}{\kappa^2} \left[ - \frac{f\left(r_2\right) - 1}{{r_2}^2} + \frac{1}{2} f''\left(r_2\right)\right]
= \frac{\e^{2\nu(r_2)}}{\kappa^2} \left[ - \frac{1 - \frac{2M}{r_2} - 1}{{r_2}^2} + \frac{1}{2} \left( - \frac{4M}{{r_2}^3} \right)\right] =0 \, , \quad
B\left(r_2\right) = 0 \, , \nonumber \\
C\left(r_2\right) =&\, \frac{\e^{-2\nu\left(r_2\right)}}{\kappa^2} \left[ \frac{f\left(r_2\right) - 1}{r^2} - \frac{1}{2} f''\left(r_2\right) \right]
= \frac{\e^{-2\nu\left(r_2\right)}}{\kappa^2} \left[ \frac{1 - \frac{2M}{r_2} - 1}{{r_2}^2} - \frac{1}{2} \left( - \frac{4M}{{r_2}^3} \right)
\right] =0\, , \nonumber \\
V\left(r_2\right) =&\, \frac{1}{\kappa^2} \left[ - \frac{f'\left(r_2\right)}{r_2} - \frac{f\left(r_2\right) - 1}{{r_2}^2} \right]
= \frac{1}{\kappa^2} \left[ - \frac{\frac{2M}{{r_2}^2}}{r_2} - \frac{1 - \frac{2M}{r_2} - 1}{{r_2}^2} \right] = 0 \, .
\end{align}
Here we have used the equations (\ref{derfs}) in the calculation of the above equations.
Eqs.~(\ref{ABCVr1}) and (\ref{ABCVr2}) are, of course, the results of the continuities of $\e^{2\nu}$, $\left(\e^{2\nu}\right)'$, and $\left(\e^{2\nu}\right)''$
at the two boundaries $r=r_1$ and $r=r_2$. The behavior of Eq.~(\ref{scalar1}) are shown in Figures.~\ref{Fig:1}~\subref{fig:A}, \subref{fig:C}, and \subref{fig:V} for the case of (\ref{Poly1})
and in Figures.~\ref{Fig:1B} \subref{fig:A2}, \subref{fig:C2}, and \subref{fig:V2} for the case of (\ref{Poly2}).

\subsection{Smooth function model}

In this subsection, we try to connect the interior region and the exterior region by using smooth functions, which describe the asymptotically Schwarzschild
spacetime for large $r$
but describes the asymptotically de Sitter spacetime for small $r$.

\subsubsection{Example 1}

We now consider the following model
\begin{align}
\label{hayward1}
\e^{2\nu}=\e^{-2\mu} = 1 - \frac{2Mr^2}{r^3 + {r_0}^3} \, .
\end{align}
The spacetime described by (\ref{hayward1}) has two horizons $\e^{2\nu}=\e^{-2\mu}=0$ in general, which corresponds to the Hayward black hole \cite{Hayward:2005gi}.
If we choose, however,
\begin{align}
\label{cond2}
\frac{2^\frac{5}{3}M}{3r_0}<1\, ,
\end{align}
$\e^{2\nu}$ and $\e^{-2\mu}$ are always positive and there does not appear any horizon.
In the following, we only consider the case that Eq.~(\ref{cond2}) is satisfied.

When $r$ is small, $\e^{2\nu}=\e^{-2\mu}$ behaves as $\e^{2\nu}=\e^{-2\mu}\sim 1 - \frac{2M}{{r_0}^3}r^2$, which is the de Sitter spacetime and
we identify $\lambda^2= \frac{2M}{{r_0}^3}$.
On the other hand, when $r$ is large, we find $\e^{2\nu}=\e^{-2\mu}\sim 1 - \frac{2M} r$, which is nothing but the Schwarzschild spacetime.
Therefore the function (\ref{hayward1}) smoothly connects the exterior region, which is the asymptotically Schwarzschild spacetime, and the interior region, which is
the asymptotically de Sitter spacetime.

By using (\ref{TSBHst2}), we can find the explicit form of $A$, $B$, $C$, and $V$, as follows,
\begin{align}
\label{TSBH6st2C}
A=&\, \frac{18M {r_0}^3r^3}{\kappa^2\left(r^3 + {r_0}^3\right)^3} \left( 1 - \frac{2Mr^2}{r^3 + {r_0}^3} \right) \, , \quad
B= 0 \, , \quad
C= - \frac{18M {r_0}^3r^3}{\kappa^2\left(r^3 + {r_0}^3\right)^3} \left( 1 - \frac{2Mr^2}{r^3 + {r_0}^3} \right)^{-1} \, , \nonumber \\
V=&\, \frac{6M{r_0}^3}{\kappa^2\left(r^3 + {r_0}^3\right)^2} \, .
\end{align}
We should note $A$ is always positive but $C$ is negative.
Therefore $\chi$ becomes a ghost if we do not add the constraint terms (\ref{lambda1}).
Due to the constraint terms (\ref{lambda1}), both of $\phi$ and $\chi$ become non-dynamical and the ghosts can be eliminated.
Because $\phi$ is canonical although $\chi$ is not canonical, we may impose the constraint on only $\chi$ by using
\begin{align}
\label{lambda1chi}
S_{\lambda_\chi} = \int d^4 x \sqrt{-g} \lambda_\chi \left( \e^{-2\mu(r=\chi)} \partial_\mu \chi \partial^\mu \chi - 1 \right) \, .
\end{align}
Instead of Eq.~(\ref{lambda1}) so that only $\chi$ is not dynamical.

\subsubsection{Example 2}

\begin{figure}
\centering
\subfigure[~The behavior of the function $\e^{2\nu}$ given by Eq.~(\ref{sad}) ]{\label{fig:ft}\includegraphics[scale=0.3]{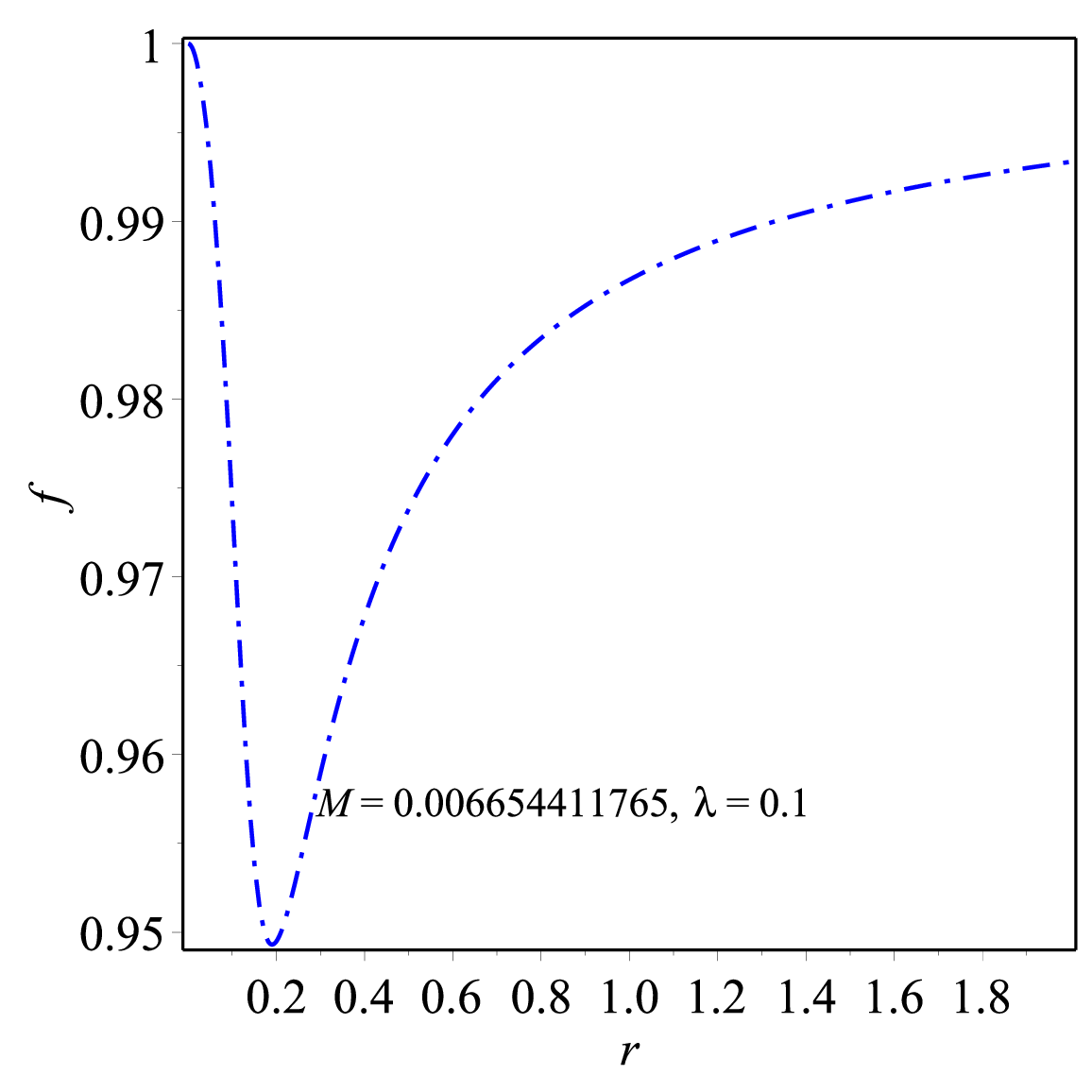}}
\subfigure[~The behavior of the scalar field $A(r)$ given by Eqs.~(\ref{ABCVex2}) ]{\label{fig:At}\includegraphics[scale=0.3]{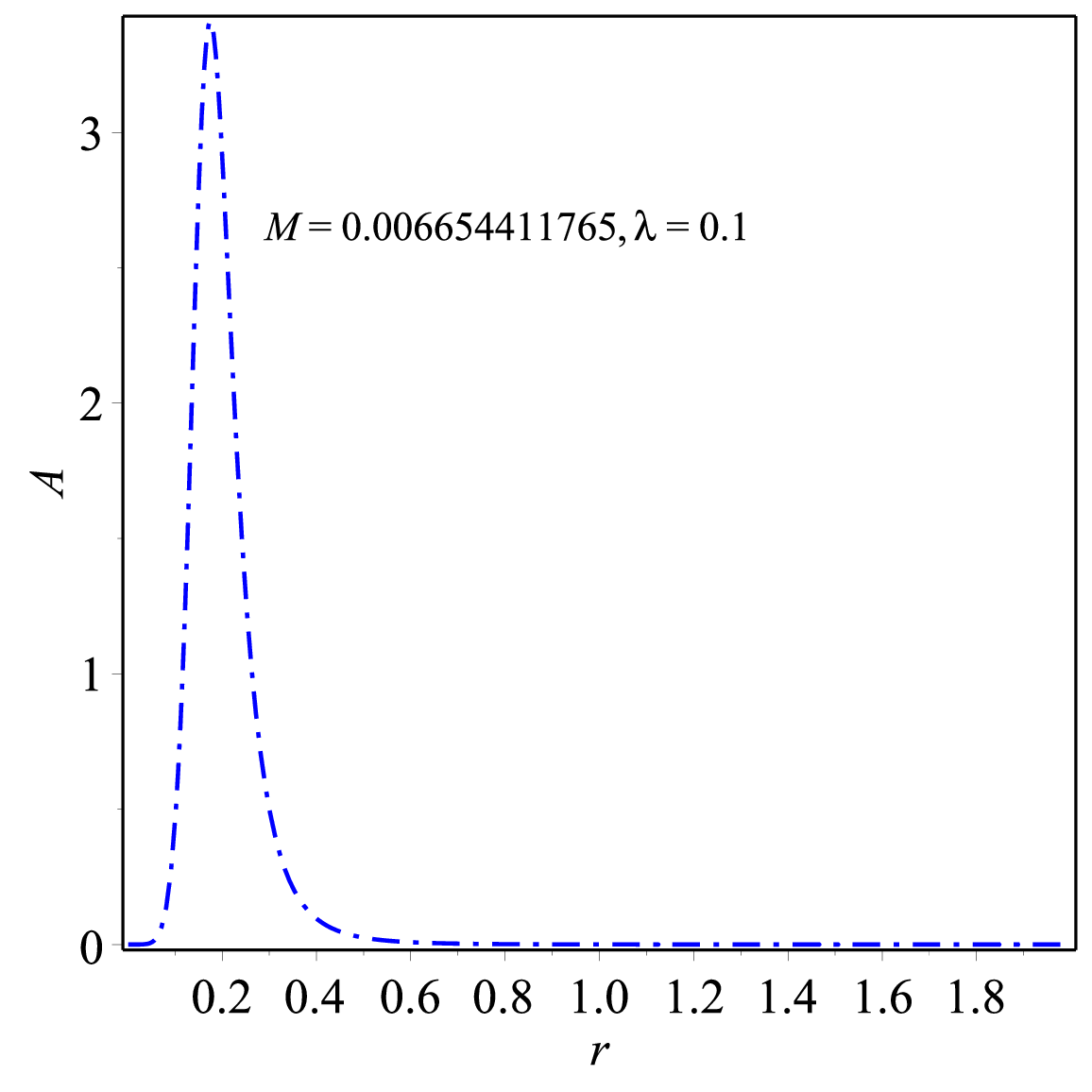}}
\subfigure[~The behavior of the scalar field $C(r)$ given by Eqs.~(\ref{ABCVex2}) ]{\label{fig:Ct}\includegraphics[scale=0.3]{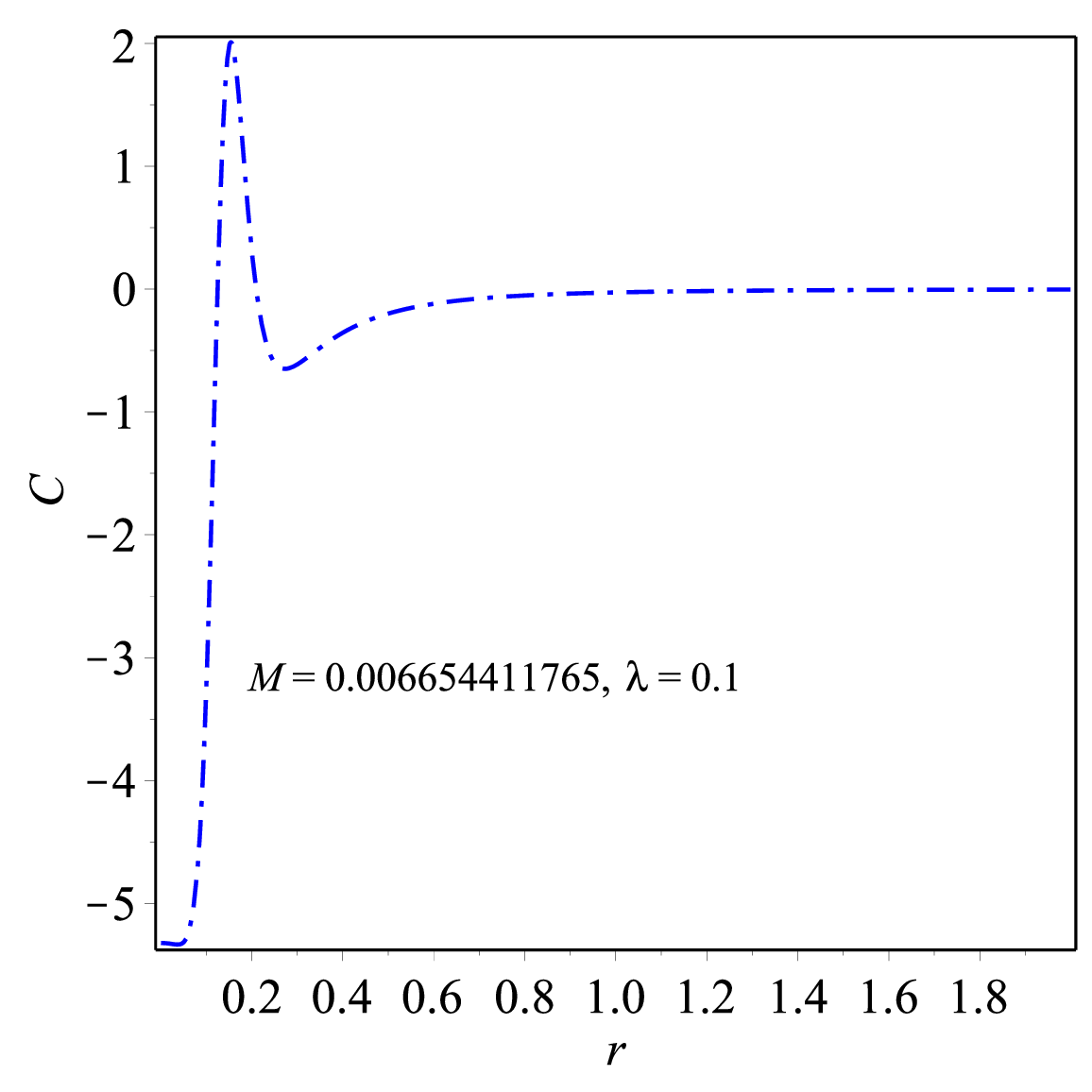}}
\subfigure[~The behavior of the potential field $V(r)$ given by Eqs.~(\ref{ABCVex2}) ]{\label{fig:Vt}\includegraphics[scale=0.3]{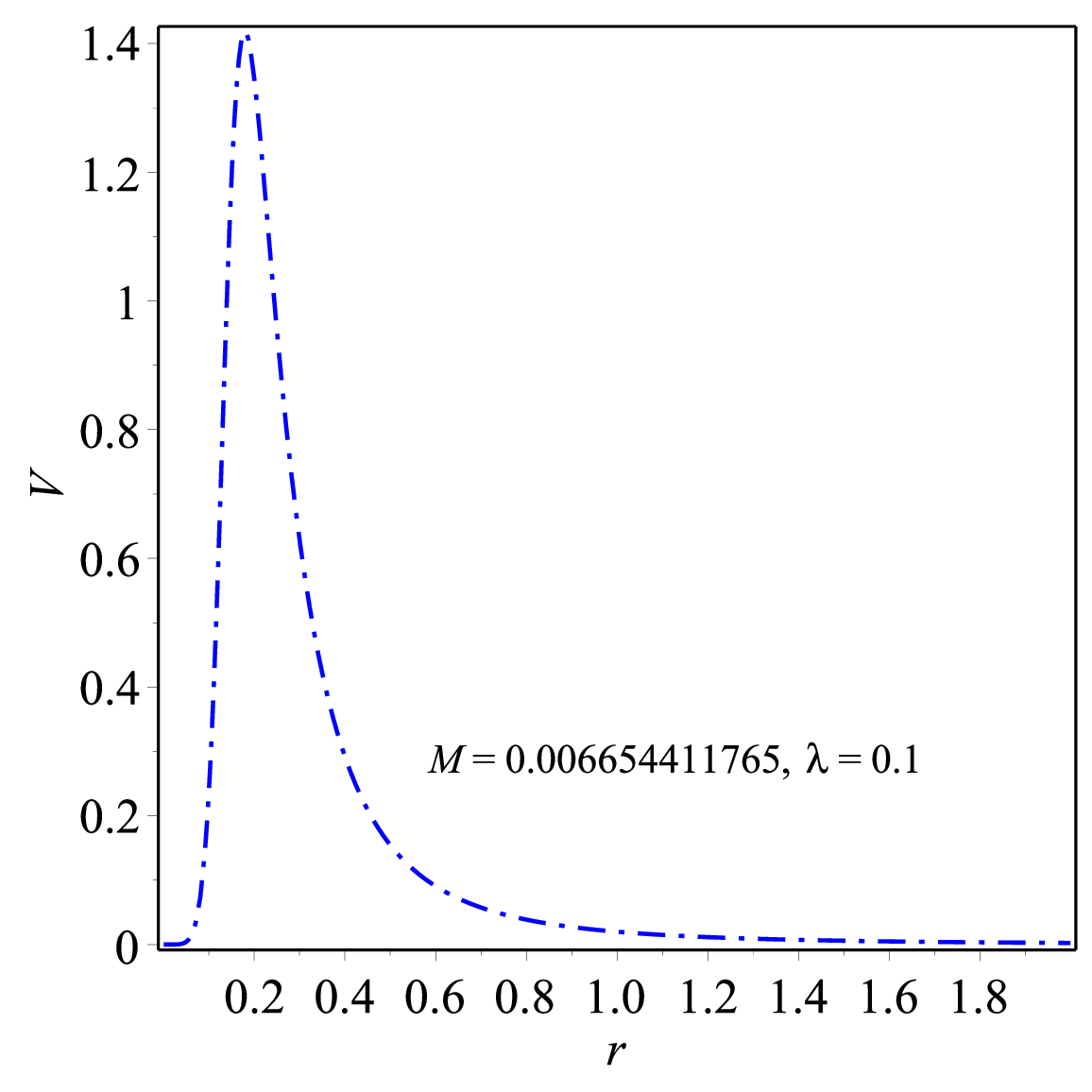}}
\caption[figtopcap]{\small{Plots of Fig. \subref{fig:ft} the function given by Eq.~(\ref{sad}); Figures~\subref{fig:At} \subref{fig:Ct}, and \subref{fig:Vt} show the behavior
of the scalars $A(r)$, $C(r)$, and the potential $V(r)$ given by Eq.~(\ref{ABCVex2}).}}
\label{Fig:2}
\end{figure}

Now let us use the following function
\begin{align}
\label{sad}
\e^{2\nu(r)}=\e^{-2\mu(r)}=1-\frac{4M\arctan\left(\frac{\pi r^3}{L^3}\right)}{\pi r}\,.
\end{align}
When $r$ is large, we find that the spacetime becomes the asymptotically Schwarzschild spacetime $\e^{2\nu(r)}=\e^{-2\mu(r)} \to 1 - \frac{2M} r$
and when $r$ is small, we obtain the de Sitter spacetime $\e^{2\nu(r)}=\e^{-2\mu(r)} \to 1 - \frac{4\pi M}{L^3} r^2$ by identifying $\lambda^2 = \frac{4\pi M}{L^3}$.
The behavior of Eq.~(\ref{sad}) is shown in Figure~\ref{Fig:2} \subref{fig:ft}.

Using Eq.~(\ref{sad}) in Eq.~(\ref{TSBHst2}), we obtain
\begin{align}
\label{ABCVex2}
A=&\, \frac{36M r^6L^3\pi^2}{\kappa^2\left( L^6+ r^6\pi^2 \right)^2}\left(1-\frac{4M\arctan\left( \frac{ r^3\pi}{L^3}\right)}{\pi r}\right) \,,\nonumber \\
C=&\, \frac{4M\left\{ 2(L^6+ r^6\pi^2)^2\arctan\left( \frac{ r^3\pi}{L^3}\right)-9 r^9L^3\pi^3\right\}}
{\kappa^2 r^2\left( L^6+ r^6\pi^2 \right)^2\left[4M\arctan\left( \frac{ r^3\pi}{L^3}\right)-\pi r\right]}\,,\nonumber \\
V=&\, \frac{6M\left\{\left( L^6+ r^6\pi^2 \right)\arctan\left( \frac{ r^3\pi}{L^3}\right)- r^3L^3\pi\right\}}
{\kappa^2 r^3\pi \left( L^6+ r^6\pi^2 \right)}\, .
\end{align}
Again, we find $A$ is always positive but $C$ is negative.
Therefore $\chi$ becomes a ghost and we need the constraint terms in (\ref{lambda1chi}).
The behavior of Eq.~(\ref{ABCVex2}) are shown in Figures~\ref{Fig:2} \subref{fig:At}, \subref{fig:Ct}, and \subref{fig:Vt} for $M=0.006654411765$ and
$\lambda = \sqrt{\frac{4\pi M}{L^3}}=0.1$.

\section{The physical characteristic of the model}\label{S4}

We assume the shell region that connects the interior region and the exterior region is rather thin and therefore the matter density composed of the scalar fiels could be high.
We now investigate the proper thickness and the surface redshift.

\subsection{Proper length of shell}

In accordance with the proposal of Mazur and Mottola, the incompressible fluid within the shell is located at the intersection of two separate regions.
The shell spans from radius $r_1$ (indicating the surface between the insdie and the shell) to radius $r_2$ (indicating the surface between the shell and the external spacetime),
with the assumption that $\lambda\left( r_2 - r_1 \right) \ll 1$.
This allows for the calculation of the proper spacing separating these two interfaces using the following equation:
\begin{align}
\label{pro}
l=\int_{r_1}^{r_2} \e^{\mu(r)} dr\,.
\end{align}
In this subsection, we only consider the case that the shell is given by the polynomial in (\ref{fr}) because the boundaries of the shell are well defined to be $r=r_1$ and $r=r_2$.

By using Eq.~(\ref{fr}) in Eq.~(\ref{pro}), we obtain,
\begin{align}
\label{frp}
l\approx&\, \frac{r_2 - r_1}{\mathcal{N}} -\frac{3 \lambda^2 \left( {r_2}^2-{r_1}^2 \right) r_2 {r_1}^2 \left( -15{r_2}^3 + {r_2}^2 r_1 + r_2 {r_1}^2 + {r_1}^3 \right)}{4\mathcal{N} \mathcal M} \nonumber \\
&\, +\frac{\left( {r_2}^3-{r_1}^3 \right)}{3\mathcal{N}} \left( \frac{\lambda^2 \left(-10{r_2}^5 -10 {r_2}^4 r_1 - 55 {r_2}^3{r_1}^2 +19 {r_2}^2{r_1}^3 + r_2 {r_1}^4 + {r_1}^5 \right) }{2\mathcal M}
\right. \nonumber \\
&\, \left. + \frac{27\lambda^4{r_2}^2{r_1}^4 \left( -15 {r_2}^3 +{r_2}^2 r_1 + r_2 {r_1}^2 + {r_1}^3 \right)^2}{8\mathcal M^2} \right)
+ \frac{\left({r_2}^4-{r_1}^4 \right)}{4\mathcal{N}} \left( - \frac{3\lambda^2r_2 \left( -5 {r_2}^3 - 5 {r_2}^2 r_1 - 5 r_2 {r_1}^2 + 3 {r_1}^3\right) }{2\mathcal M} \right. \nonumber \\
&\, - \frac{9\lambda^4r_2 {r_1}^2 \left( -15 {r_2}^3 +{r_2}^2r_1 + r_2 {r_1}^2 + {r_1}^3 \right) \left(-10 {r_2}^5 -10 {r_2}^4 r_1- 55 {r_2}^3 {r_1}^2 + 19 {r_2}^2 {r_1}^3 + r_2 {r_1}^4 + {r_1}^5 \right)}
{4 \mathcal M^2} \nonumber \\
&\, \left. - \frac{135\lambda^6{r_2}^3{r_1}^6 \left( -15 {r_2}^3 + {r_2}^2 r_1 + r_2 {r_1}^2 + {r_1}^3 \right)^3}{16 \mathcal M^3} \right)
+\frac{\left( {r_2}^5-{r_1}^5 \right)}{5\mathcal{N}}\left( \frac{3\lambda^2r_2\, \left( -2 {r_2}^2 - 2 r_2 r_1+ {r_1}^2 \right) }{2\mathcal M} \right. \nonumber \\
&\, + \frac{27\lambda^4{r_2}^2{r_1}^2 \left( -15 {r_2}^3 + {r_2}^2r_1 + r_2 {r_1}^2 +{r_1}^3 \right) \left( -5 {r_2}^3 - 5 {r_2}^2 r_1 -5 r_2 {r_1}^2 + 3{r_1}^3 \right) }{4\mathcal M^2} \nonumber \\
&\, + \frac{3\lambda^4 \left( -10 {r_2}^5 - 10 {r_2}^4 r_1 - 55 {r_2}^3 {r_1}^2 + 19 {r_2}^2 {r_1}^3 + r_2 {r_1}^4 + {r_1}^5\right)^2}{8 \mathcal M^2} \nonumber \\
&\, + \frac{135 \lambda^6{r_2}^2{r_1}^4 \left( -15 {r_2}^3 +{r_2}^2 r_1 + r_2 {r_1}^2 + {r_1}^3 \right)^2 \left( -10 {r_2}^5 -10 {r_2}^4 r_1 - 55 {r_2}^3{r_1}^2 + 19 {r_2}^2 {r_1}^3 + r_2 {r_1}^4 + {r_1}^5 \right)}
{16 \mathcal M^3} \nonumber \\
&\, \left.+ \frac{2835 \lambda^8 {r_2}^4 {r_1}^{8} \left( -15{r_2}^3 +{r_2}^2r_1 + r_2 {r_1}^2 + {r_1}^3 \right)^4}{128 \mathcal M^4} \right) \, .
\end{align}
Here
\begin{align}
\label{mathcalNM}
\mathcal{N}\equiv &\, \sqrt { - \frac{\mathcal M}{ \left( 10 {r_2}^2 - 5 r_2 r_1 + {r_1}^2 \right) \left( r_2 - r_1 \right)^3}} \, , \nonumber \\
\mathcal M\equiv &\, \lambda^2 {r_2}^2 {r_1}^3 \left( 15 {r_2}^2 - 3 r_2 r_1 - 3 {r_1}^2 \right) -10 {r_2}^5 +35 {r_2}^4r_1 - 46 {r_2}^3 {r_1}^2 + 28 {r_2}^2 {r_1}^3 - 8 r_2 {r_1}^4 - {r_1}^5 \, .
\end{align}
For example, in the case of (\ref{Poly1}), we find $\lambda l=0.0201\cdots$, which is not so changed from $\lambda \left( r_2 - r_1 \right) = 0.02$.

\subsection{Surface redshift}

In this particular subsection, we will examine the surface redshift.
The examination of gravastar surface redshift stands as a crucial means for assessing its detection and the surface redshift has been often used to investigate the stability
of the gravastar configuration.
Surface gravitational redshift, denoted as $Z_s$, is defined by using the wavelength
$\lambda_e$ of the emitted signal at the surface of the gravastar
and the wavelength $\lambda_0$ of the observed signal, that is, $Z_s = \frac{\lambda_0 - \lambda_e}{\lambda_e}$.

Buchdahl's claim, founded on the conditions of isotropy, staticity, and a perfect fluid distribution, suggests that the surface redshift should remain below 2,
denoted as $Z_s < 2$ \cite{Buchdahl:1959zz, Straumann:1984xf, Boehmer:2006ye}.
Conversely, Ivanov \cite{Ivanov:2002xf} argued that in the case of an anisotropic fluid distribution, it could increase to as high as 3.84.
Furthermore, Barraco and Hamity \cite{Barraco:2002ds} showed that in the absence of a cosmological constant, for an isotropic fluid distribution, $Z_s$
stays at a value less than or equal to 2.
Nevertheless, Bohmer and Harko \cite{Boehmer:2006ye} demonstrated that in an anisotropic star with the influence of a cosmological constant, $Z_s$ can reach as high as 5. 
The above limits have been obtained for compact objects that satisfy
some suitable energy conditions. Therefore, if we violate the energy conditions, we can construct objects that do not satisfy the above conditions for redshift. 
Even in our model, the energy conditions are not satisfied.
The difference is that the objects are assumed to be given by perfect fluid in the previous works but
as we have shown, the scalar fields in this paper cannot be approximated to be perfect fluid. In fact, in the case of the fluid, the sound wave 
is produced by the compression and expansion of the fluid. 
If the energy conditions are violated, such an oscillation generates instability. In our model, however, as we have shown after Eq.~(\ref{lambda2}), 
there appear to be no oscillations in the perturbation of the scalar fields. 
Therefore, the sound wave does not appear, and the associated instability is not generated. 
The constraints outlined in (\ref{lambda2}) were employed to emulate the behavior of dark matter, as described in \cite{Chamseddine:2014vna}.
Once more, it's essential to emphasize that effective dark matter is not actual dark matter, and it does not undergo gravitational collapse.
In the model presented in this paper, the arrangement remains stable when subjected to perturbations.
To compare with earlier studies employing conventional fluids, we assess the surface redshift within the current model.
To calculate the surface redshift, we utilize the subsequent equation:
\begin{align}
\label{red}
Z_s=-1 + \frac{1}{\sqrt{\left| g_{tt} \right|}}\,,
\end{align}
which is evaluated at the surface of the gravastar.
Therefore in the polynomial model (\ref{fr}), we find
\begin{align}
\label{redpln}
Z_s=-1 + \frac{1}{\sqrt{1 - \frac{2M}{r_2}}}\,,
\end{align}
because the surface is given by $r=r_2$.
In the case of small $\lambda$ in (\ref{Poly1}), we obtain
\begin{align}
\label{redpln2}
Z_s \sim \frac M{r_2} \fallingdotseq 0.0054\cdots\, ,
\end{align}
which is very small and therefore the value is in the stable region.
On the other hand, in the case of thin shell in (\ref{Poly2}), we find,
\begin{align}
\label{redpln3}
Z_s =9\, ,
\end{align}
which is beyond the stable region in the previous works \cite{Buchdahl:1959zz, Straumann:1984xf, Boehmer:2006ye, Ivanov:2002xf, Barraco:2002ds, Boehmer:2006ye}
and in the region prohibited by the instability.
We should note, again, that the solution in the present model is always stable under fluctuations.
Therefore we can realize the stable gravastar with large surface redshift.

\begin{figure}
\centering
\includegraphics[scale=0.3]{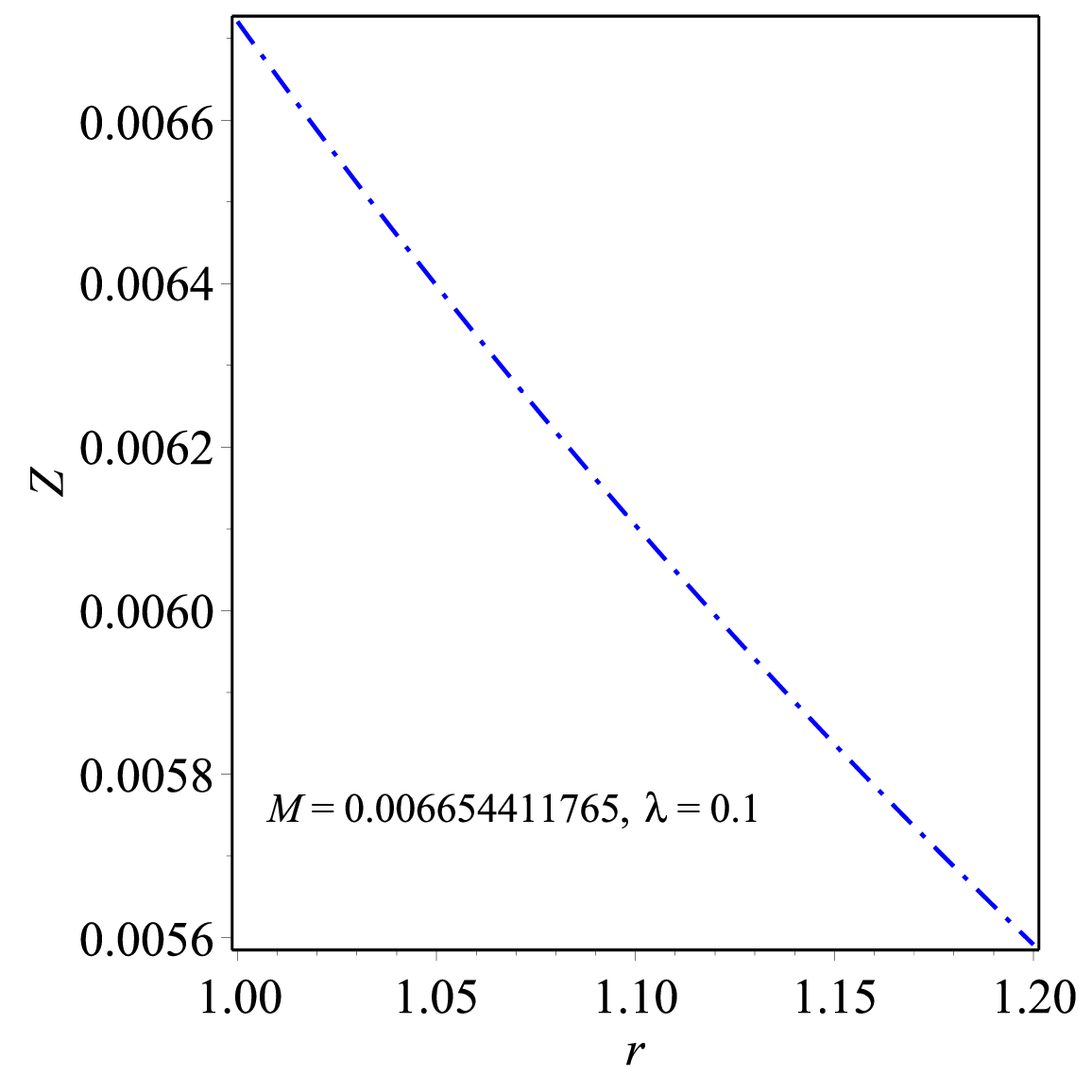}
\caption[figtopcap]{\small{Plot of the redshift given by Eq.~(\ref{red1}).}}
\label{Fig:3}
\end{figure}

In the case of smooth function models in (\ref{hayward1}) and (\ref{sad}),
the position of the surface is not well-defined.
We now evaluate the redshift in some region of $r$.
In the case of (\ref{hayward1}), by using Eq.~(\ref{red}) for general $r$, we obtain
\begin{align}
\label{red1}
Z_s=-1+\frac{1}{\sqrt{\frac{2Mr^2}{r^3 + {r_0}^3}}}\,,
\end{align}
and for Eq.~(\ref{sad}), we obtain the following redshift as,
\begin{align}
\label{red2}
Z_s=-1+\frac{1}{\sqrt{1-\frac{4M\arctan\left(\frac{\pi r^3}{L^3}\right)}{\pi r}}}\,.
\end{align}
The behavior of Eq.~(\ref{red1}) is shown in Figure~\ref{Fig:3}
for $M=0.006654411765$ and $\lambda = \sqrt{\frac{4\pi M}{L^3}}=0.1$, again,
which ensures that the redshift in our model is in the stable region given
in \cite{Buchdahl:1959zz, Straumann:1984xf, Boehmer:2006ye}.
The behavior of redshift given by Eq.~(\ref{red2}) has the same pattern as Eq.~(\ref{red1}) which tells us that the redshift is in the stable region.

\section{Effective equation of state parameter}

The matter that composes the shell of the gravastar obeys a rather excentric equation of state (EoS).
In our model, although the shell is composed of scalar fields, we investigate what kind of matter corresponds to the shell in the polynomial model (\ref{fr}).

From Eqs.~(\ref{TSBH2}), (\ref{TSBH3}), and (\ref{TSBH4}), one can find the energy density $\rho$, radial and tangential pressures $p_r$ and $p_t$, as follows,
\begin{align}
\label{rhop}
\rho=&\, \frac{\kappa^2}{2} \left[ 2 \e^{2\nu \left( r \right)}V \left( r \right) +A \left( r \right) +C \left( r \right) \e^{2\nu \left( r \right) -2\mu \left( r \right)} \right]\,,
\nonumber \\
p_r=&\, \frac{\kappa^2}{2} \left[C \left( r \right) +A \left( r \right) \e^{2\mu \left( r \right) -2\nu \left( r \right)} -2 \e^{2\mu \left( r \right)}
V \left( r \right) \right] \,, \nonumber \\
p_t=&\, \frac{\kappa^2}{2} \left[A \left( r \right) \e^{2\mu \left( r \right) -2\nu \left( r \right)} - 2 \e^{2\mu \left( r \right)}
V \left( r \right) -C \left( r \right) \right] \,.
\end{align}
We use the above expressions for the models proposed in this paper.

\begin{figure}
\centering
\subfigure[~The behavior of the radial EoS given by Eq.~(\ref{EoS}) ]{\label{fig:EoSr}\includegraphics[scale=0.3]{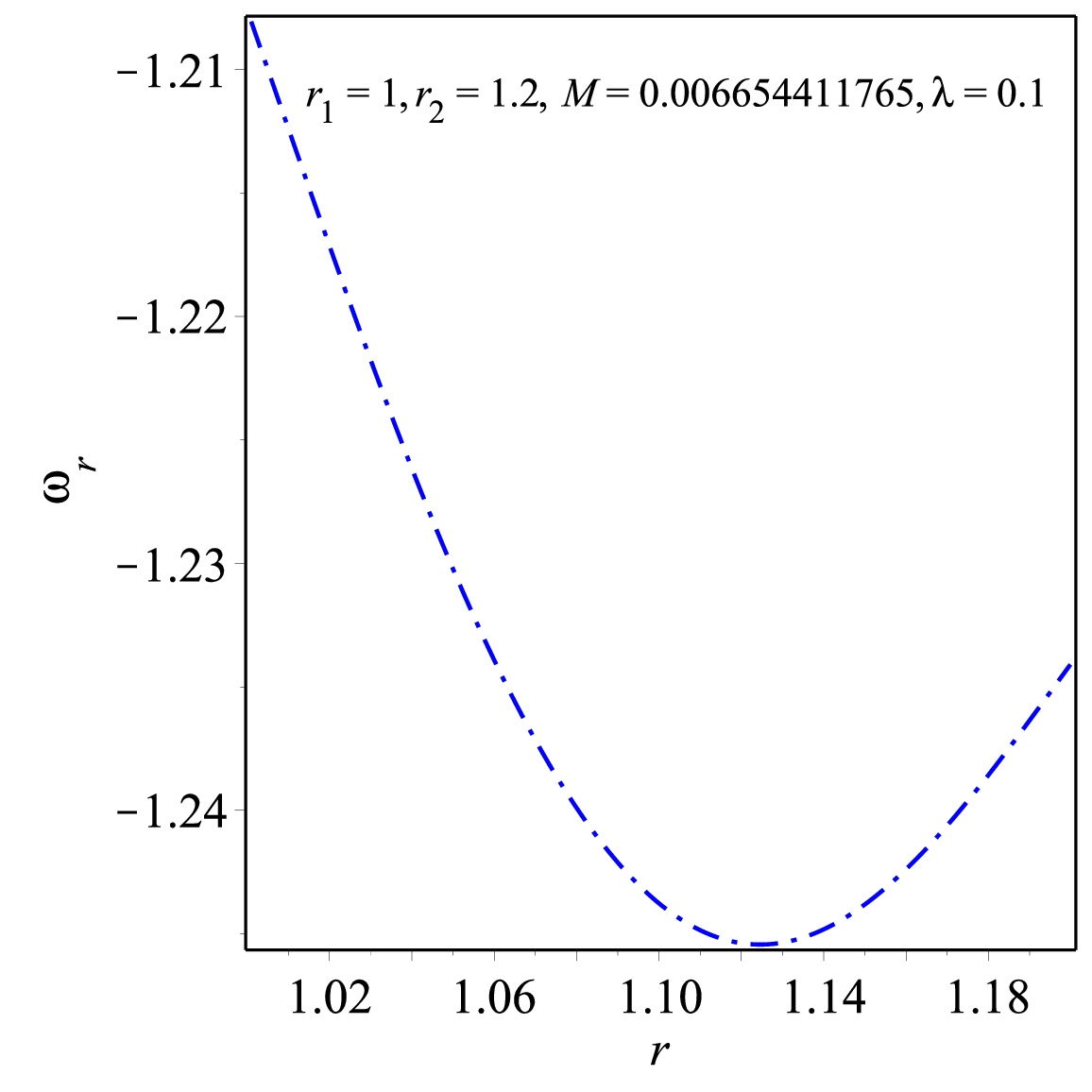}}
\subfigure[~The behavior of the tangential EoS given by Eq.~(\ref{EoS})]{\label{fig:EoSt}\includegraphics[scale=0.3]{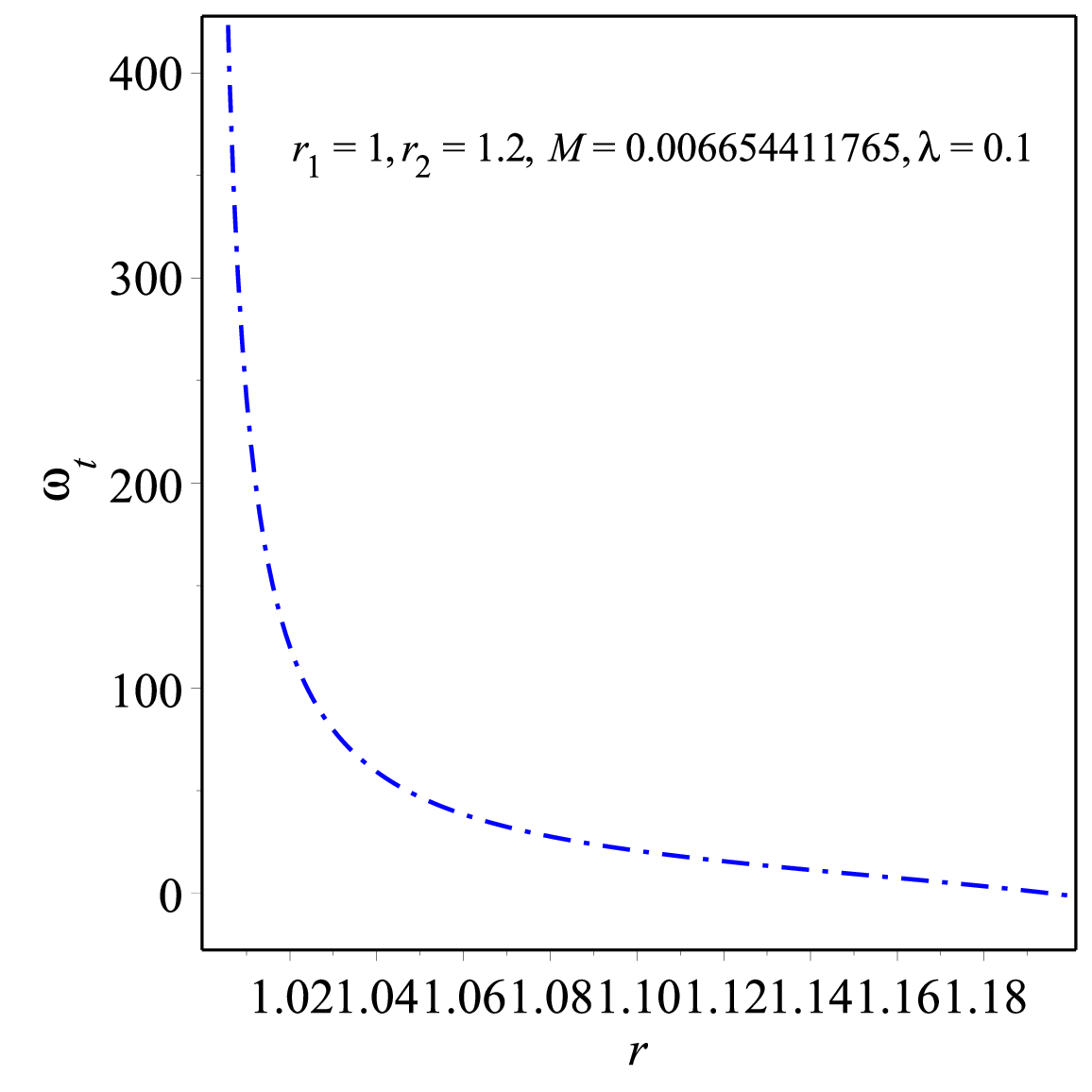}}
\caption[figtopcap]{\small{Plots of Figures for sufficiently small $\lambda$ case in (\ref{Poly1}),
\subref{fig:EoSr} the radial EoS given by Eq.~(\ref{EoS}); \subref{fig:EoSt} the tangential EoS given by Eq.~(\ref{EoS}).}}
\label{Fig:4}
\end{figure}

\begin{figure}
\centering
\subfigure[~The behavior of the radial EoS given by Eq.~(\ref{EoS}) ]{\label{fig:EoSrthin}\includegraphics[scale=0.3]{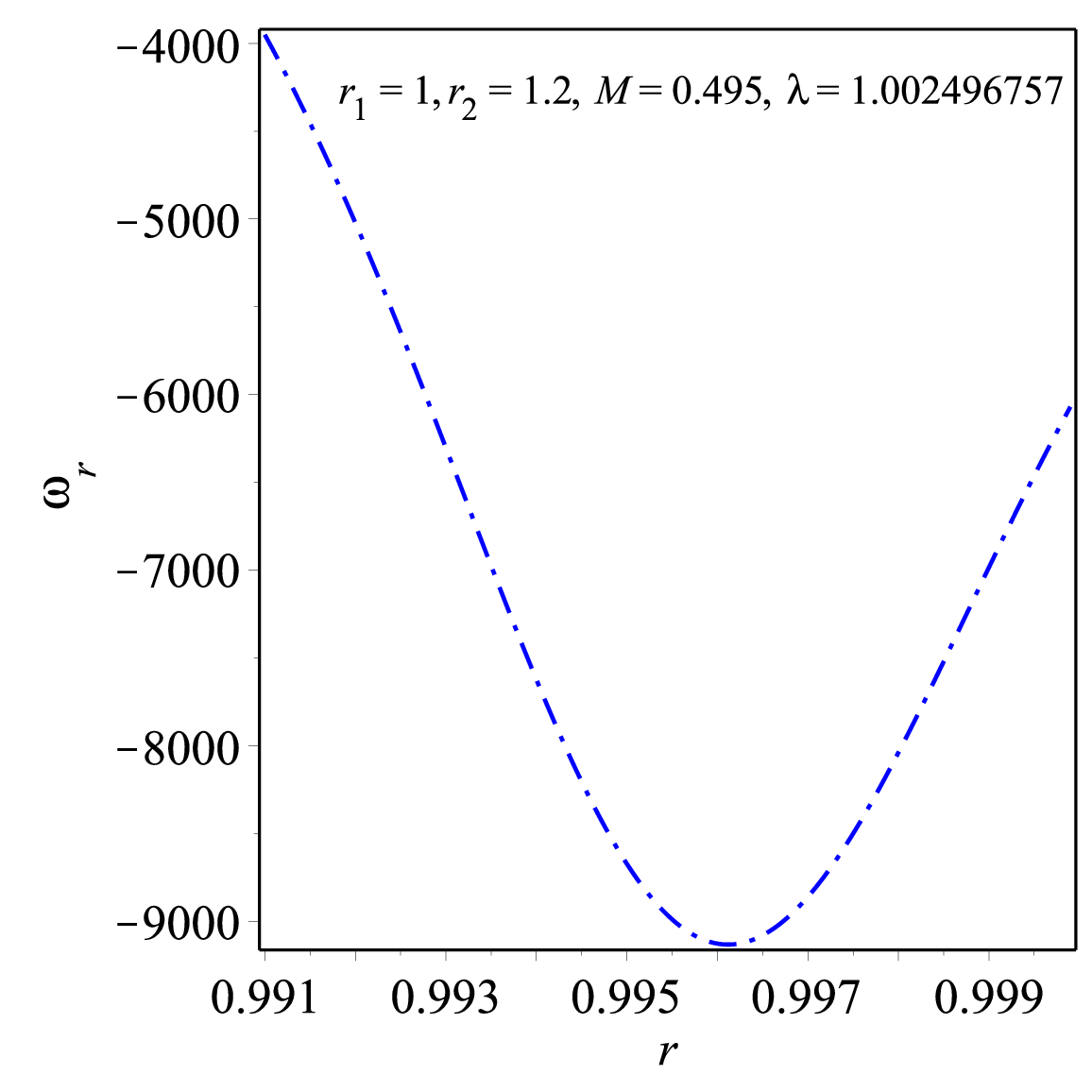}}
\subfigure[~The behavior of the tangential EoS given by Eq.~(\ref{EoS})]{\label{fig:EoStthin}\includegraphics[scale=0.3]{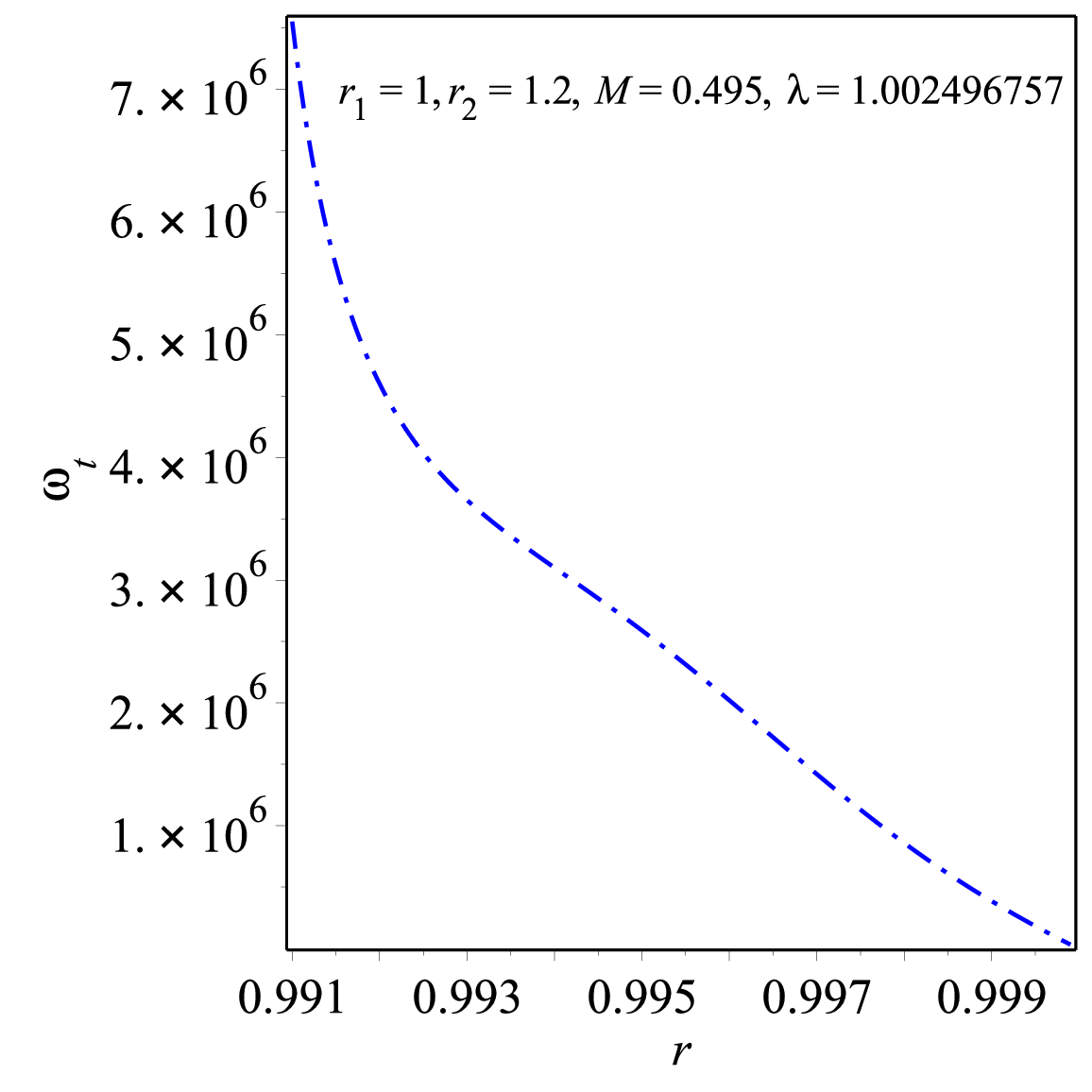}}
\caption[figtopcap]{\small{Plots of Figures for sufficiently thin shell case in (\ref{Poly2}),
\subref{fig:EoSr} the radial EoS given by Eq.~(\ref{EoS}); \subref{fig:EoSt} the tangential EoS given by Eq.~(\ref{EoS}).}}
\label{Fig:4B}
\end{figure}

In the case that the shell is described by the polynomial function (\ref{fr}), by substituting (\ref{fr}) with (\ref{as1}) and Eq.~(\ref{scalar1}),
we obtain the radial and tangential EoS parameters $\omega_r$ and $\omega_t$ as follows,
\begin{align}
\label{EoS}
\omega_r\equiv&\, \frac{p_r}{\rho}
= - \left( 10 {r_2}^2 - 5 r_2 r_1 + {r_1}^2 \right)^2 \left( r_2 - r_1 \right)^6 \left[ 10 {r_2}^5 -35 {r_2}^4 r_1 + 46 {r_2}^3 {r_1}^2 - 28 {r_2}^2 {r_1}^3 + 8 r_2 {r_1}^4-{r_1}^5 \right. \nonumber \\
&\, +\lambda^2 \left\{ \left( - 6 {r_2}^3 - 6 {r_2}^2r_1+ 3 r_2 {r_1}^2 \right) r^4
+ \left( 15 {r_2}^4 + 15 {r_2}^3 r_1 + 15 {r_2}^2 {r_1}^2 - 9 r_2 {r_1}^3 \right) r^3 \right. \nonumber \\
&\, + \left( -10 {r_2}^5- 10 {r_2}^4 r_1 - 55 {r_2}^3 {r_1}^2 + 19 {r_2}^2 {r_1}^3 + r_2 {r_1}^4 + {r_1}^5 \right) r^2
+ \left( + 45 {r_2}^4 {r_1}^2 - 3 {r_2}^3 {r_1}^3 -3 {r_2}^2 {r_1}^4 - 3 r_2 {r_1}^5 \right) r \nonumber \\
&\, \left. \left. -15 {r_2}^4 {r_1}^3 + 3 {r_2}^3 {r_1}^4 + 3 {r_2}^2 {r_1}^5 \right\} \right]^{-2}\,,\nonumber \\
\omega_t\equiv&\, \frac{p_t}{\rho} = r \left( r_2 - r_1 \right)^6 \left( 10 {r_2}^2 - 5 r_2 r_1 + {r_1}^2 \right)^2
\left\{ \left( -16 {r_2}^2 - 16r_2 r_1 + 8 {r_1}^2 \right) r^2 + \left( 15 {r_2}^3 - {r_2}^2 r_1 - r_2 {r_1}^2 - {r_1}^3 \right) r \right. \nonumber \\
&\left. + 15 {r_2}^3 r_1 - {r_2}^2 {r_1}^2 - r_2 {r_1}^3 - {r_1}^4 \right\} \left( r-r_1 \right)^{-1}
\left\{ \left( -4 {r_2}^2 - 4 r_2 r_1 + 2 {r_1}^2 \right) r^2
+ \left( 5 {r_2}^3 - 3 {r_2}^2 r_1 - 3 r_2 {r_1}^2 +r{r_1}^3 \right) r \right. \nonumber \\
&\, \left. + 10 {r_2}^3 r_1 - 2 {r_2}^2 {r_1}^2 -2 r_2 {r_1}^3 \right\}^{-1}
\left[ 10 {r_2}^5 -35 {r_2}^4 r_1 + 46 {r_2}^3 {r_1}^2 - 28 {r_2}^2 {r_1}^3 + 8 r_2 {r_1}^4 - {r_1}^5 \right. \nonumber \\
&\, + \lambda^2 \left\{ \left( - 6 {r_2}^3 -6 {r_2}^2 r_1 + 3 r_2 {r_1}^2 \right) r^4
+ \left( 15 {r_2}^4 + 15 {r_2}^3 r_1 + 15 {r_2}^2 {r_1}^2 - 9 r_2 {r_1}^3 \right) r^3
\right. \nonumber \\
&\, + \left( -10 {r_2}^5 - 10 {r_2}^4 r_1 -55 {r_2}^3 {r_1}^2 + 19 {r_2}^2 {r_1}^3 + r_2 {r_1}^4 + {r_1}^5 \right) r^2 \nonumber \\
&\, \left. \left. + \left( 45 {r_2}^4 {r_1}^2 - 3 {r_2}^3 {r_1}^3 - 3 {r_2}^2 {r_1}^4 - 3 r_2 {r_1}^5 \right) r
 -15 {r_2}^4 {r_1}^3 + 3 {r_2}^3 {r_1}^4 + 3 {r_2}^2 {r_1}^5 \right\} \right]^{-2} \,.
\end{align}
The behaviors of Eqs.~(\ref{EoS}) are shown in Figures~\ref{Fig:4}~\subref{fig:EoSr} and \subref{fig:EoSt} for sufficiently small $\lambda$ case
in (\ref{Poly1}) and in Figures~\ref{Fig:4B}~\subref{fig:EoSrthin} and \subref{fig:EoStthin} for sufficiently thin shell case
in (\ref{Poly2}).
As clear from these Figures, the signature of the radial EoS parameter $\omega_r$ is different from that of the tangential one $\omega_t$ and
especially the value of the tangential EoS parameters are very huge.
This could tell that the gravastar geometry realized in this paper by using two scalar fields could be difficult to be generated by using ordinary fluid.

\begin{figure}
\centering
\subfigure[~The behavior of the radial EoS given by Eq.~(\ref{EoSh}) ]{\label{fig:EoSrh}\includegraphics[scale=0.3]{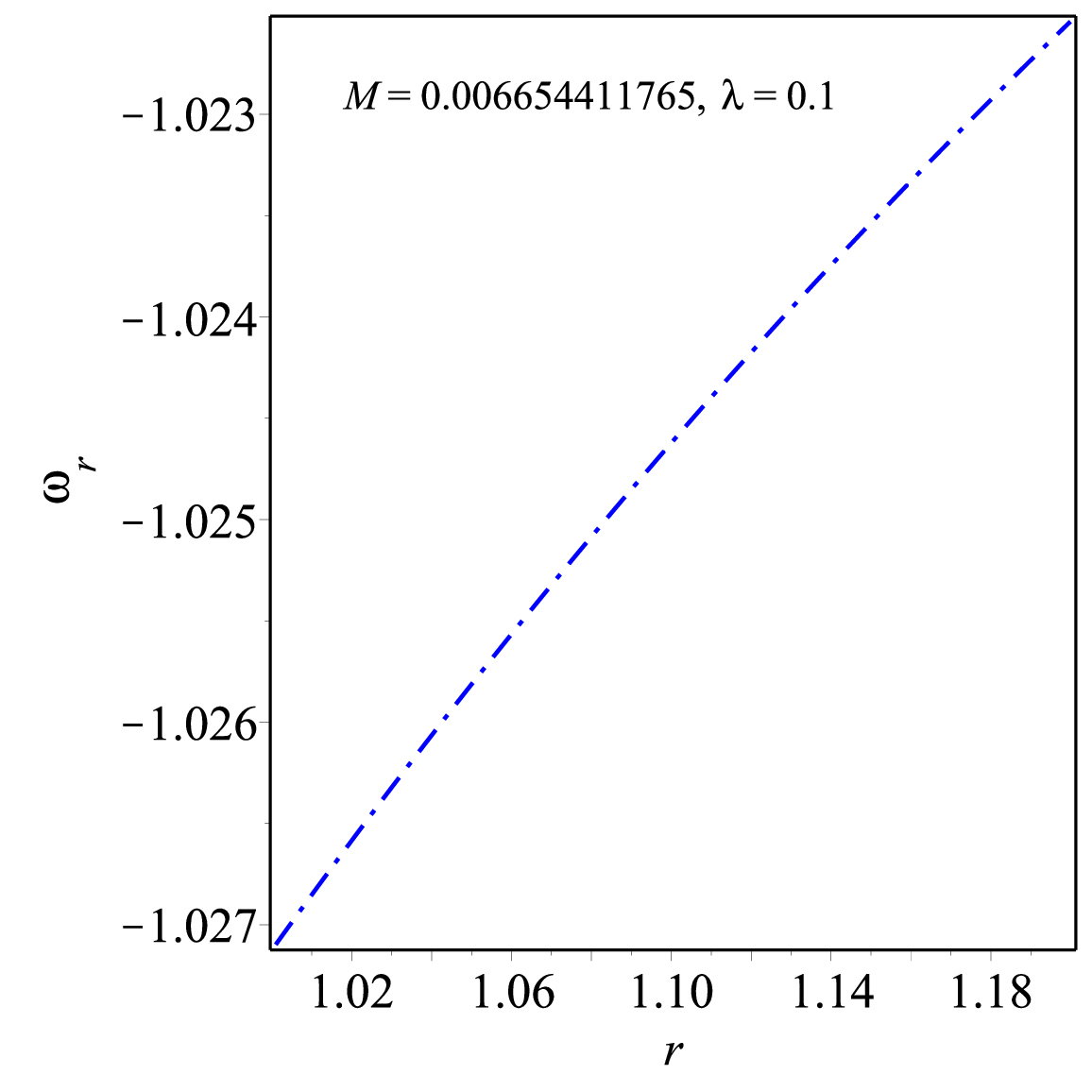}}
\subfigure[~The behavior of the tangential EoS given by Eq.~(\ref{EoSh})]{\label{fig:EoSth}\includegraphics[scale=0.3]{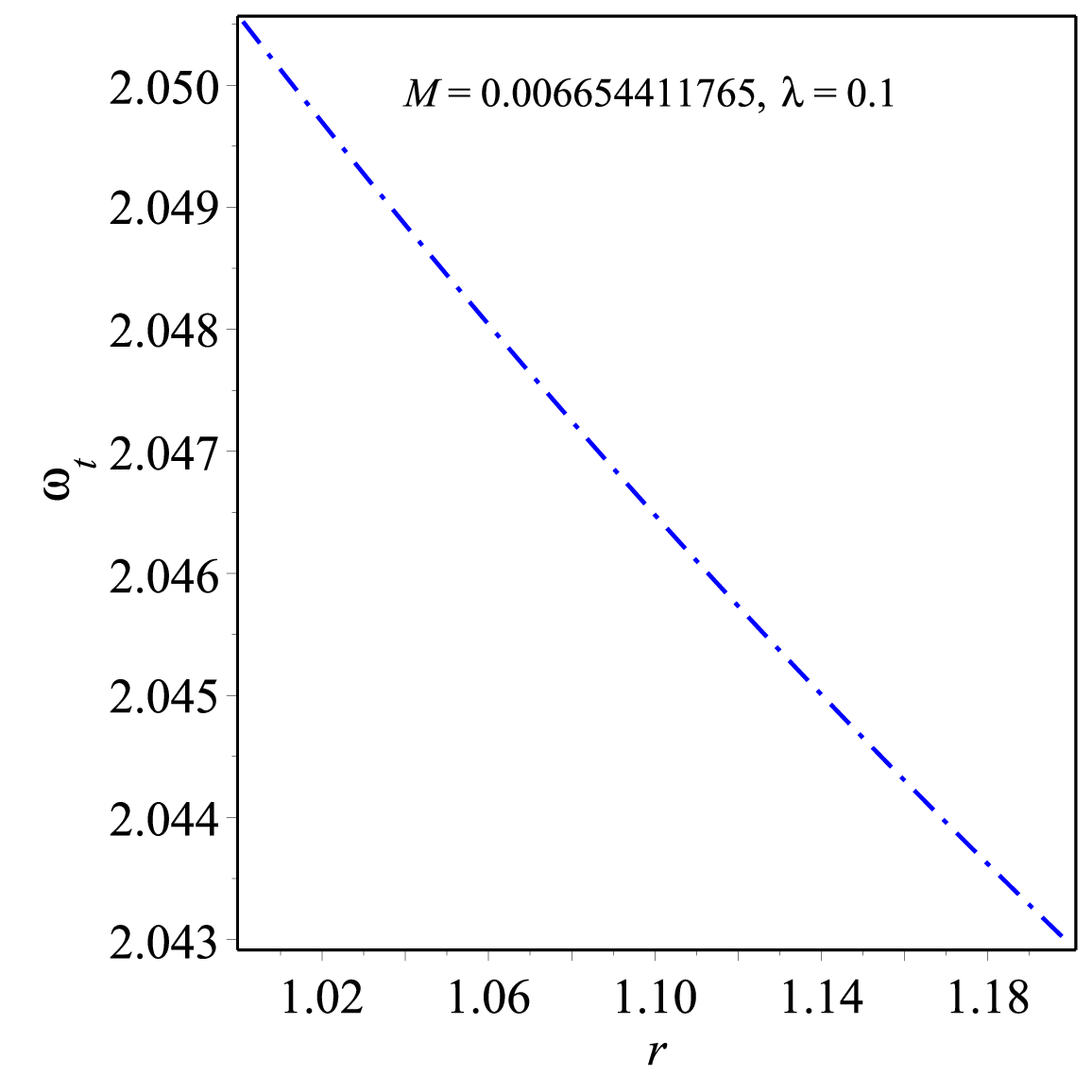}}
\caption[figtopcap]{\small{Plots of Fig. \subref{fig:EoSrh} the radial EoS given by Eq.~(\ref{EoSh}); \subref{fig:EoSth} the tangential EoS given by Eq.~(\ref{EoSh}).}}
\label{Fig:5}
\end{figure}

 From Eqs.~(\ref{hayward1}) and (\ref{TSBH6st2C}) for Example 1 in the smooth function model,
one can find the energy density $\rho$, radial and tangential pressures $p_r$ and $p_t$, as follows,
\begin{align}
\label{rhoh}
\rho=&\,- \frac{3 \left( \left( -\lambda^2+ r^2 \right) M- \frac{r^3}{2} \right) M^2\lambda^2}{ \left( M\lambda^2 + \frac{r^3}{2} \right)^3}\,,
\nonumber \\
p_r=&\, - \frac{12 \lambda^2 M^2}{ \left( r^3+2 M\lambda^2 \right) \left( r^3+2 M\lambda^2 - 2 M r^2 \right) } \,, \nonumber \\
p_t=&\, \,{\frac{3 M^2\lambda^2 \left( M\lambda^2- r^3 \right)}
{ \left( M\lambda^2 + \frac{r^3}{2} \right)^2 \left( \left( -\lambda^2+ r^2 \right) M - \frac{r^3}{2} \right) }} \,.
\end{align}
Here $\lambda = \sqrt{\frac{2M}{{r_0}^3}}$.
Using Eq.~(\ref{rhoh}), we obtain the radial and tangential EoS parameters as follows,
\begin{align}
\label{EoSh}
\omega_r = &\, \frac{ \left( 4 M\lambda^2 + \frac{r^3}{2} \right)^3}{ \left( r^3+2 M\lambda^2 \right)
\left( r^3 + 2 M\lambda^2 - 2 M r^2 \right) \left( \left( -\lambda^2 + r^2 \right) M - \frac{r^3}{2} \right) } \,,\nonumber \\
\omega_t = &\, - \frac{ \left( M\lambda^2- r^3 \right) \left( M\lambda^2 +\frac{r^3}{2} \right) }
{ \left( \left( -\lambda^2+ r^2 \right) M-\frac{r^3}{2} \right)^2} \,.
\end{align}
The behavior of Eqs.~(\ref{EoSh}) are shown in Figures~\ref{Fig:5}~\subref{fig:EoSrh} and \subref{fig:EoSth}
for $M=0.006654411765$ and $\lambda = \sqrt{\frac{2M}{{r_0}^3}}=0.1$, which satisfies the condition (\ref{cond2}).
Although the signature of the radial EoS parameter is different from that of the tangential one, the values are not surprisingly large compared
with the polynomial cases in Figures~\ref{Fig:4} and \ref{Fig:4B}.

\begin{figure}
\centering
\subfigure[~The behavior of the radial EoS given by Eq.~(\ref{EoSt}) ]{\label{fig:EoSrt}\includegraphics[scale=0.3]{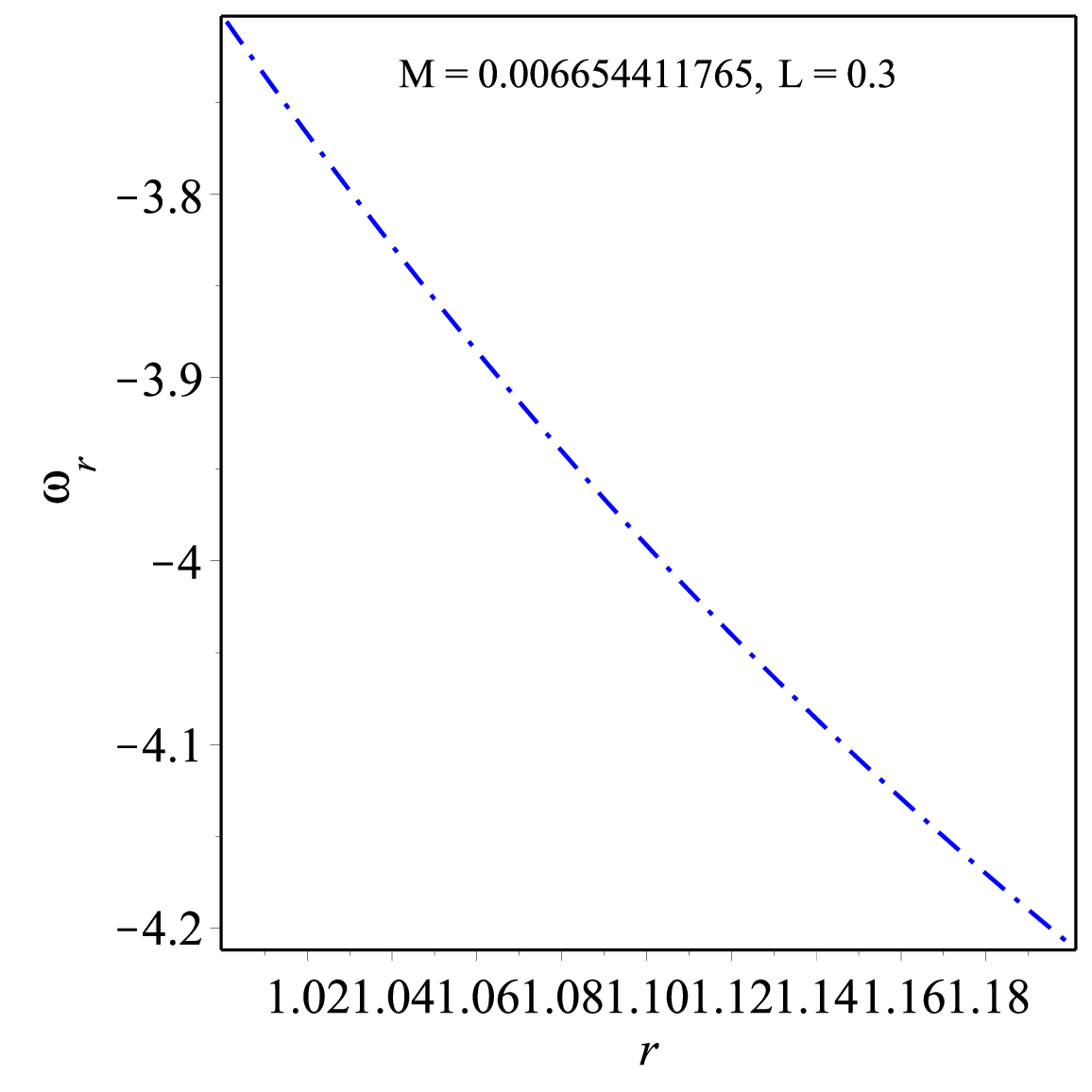}}
\subfigure[~The behavior of the tangential EoS given by Eq.~(\ref{EoSt})]{\label{fig:EoStt}\includegraphics[scale=0.3]{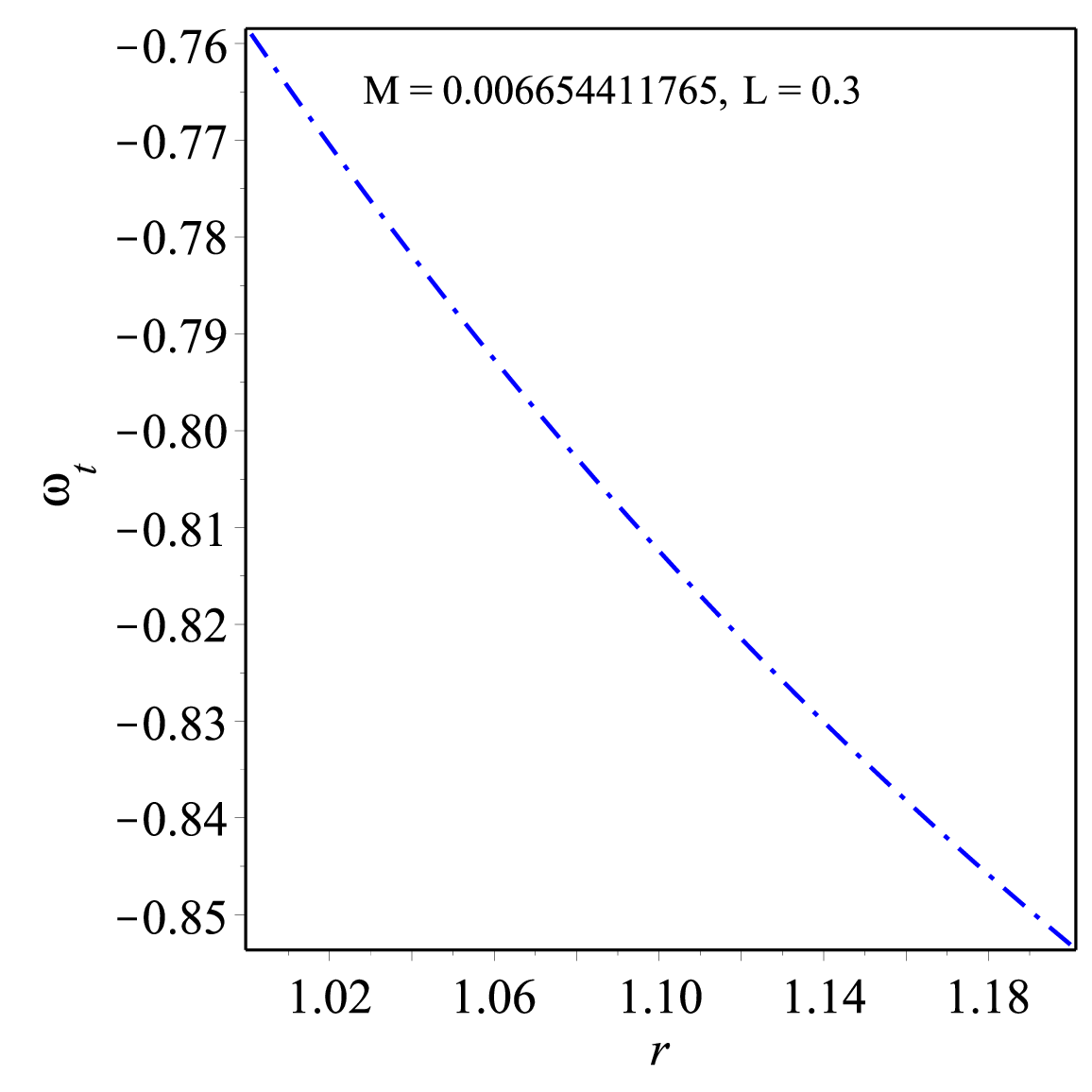}}
\caption[figtopcap]{\small{Plots of Fig. \subref{fig:EoSrt} the radial EoS given by Eq.~(\ref{EoSt}); \subref{fig:EoStt} the tangential EoS given by Eq.~(\ref{EoSt}).}}
\label{Fig:6}
\end{figure}

 From Eqs.~(\ref{sad}), and (\ref{ABCVex2}) for Example 2 in the smooth function model,
one can find the energy density $\rho$, radial and tangential pressures $p_r$, and $p_t$, as follows,
\begin{align}
\label{rhot}
\rho=&\,\frac{8M}{ r^4 \left( L^6+ r^6 \pi^2 \right)^2 \pi^2}\left( \left( L^6+ r^6 \pi^2 \right)^2
\arctan \left( {\frac{ r^3\pi }{ L^3}} \right) +15 r^9 L^3 \pi^3 - 3 r^3 L^9\pi \right)
\left( \frac{\pi r}{4} -M \arctan \left( {\frac{ r^3\pi }{ L^3}} \right) \right) \kappa^2\,,
\nonumber \\
p_r=&\,-\frac{ 10 M \kappa^2}{ r^2 \left( \pi \,r-4\,M\arctan \left( {\frac{ r^3\pi }{ L^3}} \right) \right)
\left( L^6+ r^6 \pi^2 \right)^2}\left( \left( L^6+ r^6 \pi^2 \right)^2\arctan \left( {\frac{ r^3\pi }{ L^3}} \right) - \frac{3 L^3 r^3 \pi}{5}
\left( 7 r^6 \pi^2 + L^6 \right) \right) \,, \nonumber \\
p_t=&\,\frac{2 M\kappa^2}{ \left( -\pi r + 4 M\arctan \left( \frac{ r^3\pi}{ L^3} \right) \right) r^2
\left( L^6+ r^6 \pi^2 \right)} \left( \left( L^6+ r^6 \pi^2 \right) \arctan \left( {\frac{ r^3\pi }{ L^3}} \right) -3 r^3 L^3\pi \right) \,.
\end{align}
Using Eq.~(\ref{rhoh}), we obtain the radial and tangential EoS parameters as follows,
\begin{align}
\label{EoSt}
\omega_r = &\, -\frac{\left( 5\left( L^6+ r^6 \pi^2 \right)^2\arctan \left( {\frac{ r^3\pi }{ L^3}} \right)
 -3 L^3 r^3 \pi \left( 7 r^6 \pi^2 + L^6 \right) \right) r^2 \pi^2 }{\left(\pi r - 4 M\arctan \left( {\frac{ r^3\pi }{ L^3}} \right) \right)^2
\left( \left( L^6+ r^6 \pi^2 \right)^2 \arctan \left( {\frac{ r^3\pi }{ L^3}} \right) + 15 r^9 L^3 \pi^3 - 3 r^3 L^9\pi \right)} \,,\nonumber \\
\omega_t\equiv&\,-\frac{ \left( \left( L^6+ r^6 \pi^2 \right) \arctan \left( {\frac{ r^3\pi }{ L^3}} \right)
 -3 r^3 L^3\pi \right) r^2 \left( L^6+ r^6 \pi^2 \right) \pi^2}{\left(\pi r - 4 M \arctan \left( {\frac{ r^3\pi }{ L^3}} \right) \right)^2
\left( \left( L^6 + r^6 \pi^2 \right)^2 \arctan \left( {\frac{ r^3\pi }{ L^3}} \right) +15 r^9 L^3 \pi^3 - 3 r^3 L^9\pi \right)} \,.
\end{align}
The behavior of Eqs.~(\ref{EoSt}) are shown in Figures~\ref{Fig:6}~\subref{fig:EoSrt} and \subref{fig:EoStt}
for $M=0.006654411765$ and $\lambda = \sqrt{\frac{4\pi M}{L^3}}=0.1$.
In this case, both EoS parameters are negative.

\section{Summary and discussion}\label{S5}

The intention behind the gravastar solution, incorporating a modified matter source within the framework of Einstein's GR, is to investigate alternative gravastar models
and examine the consequences they may entail.
Gravastars are theoretical entities suggested as potential substitutes for black holes.
These objects are believed to consist of exotic material capable of obstructing the creation of an event horizon, a characteristic typically associated with black holes.
In contrast, gravastars feature a surface referred to as a ``gravitational vacuum star'', which can replicate certain attributes of a black hole but lacks the singularity at its core.
In this particular scenario, the background is assumed to be Einstein's GR with two scalar fields.

Applying the spherically symmetric spacetime to the field equation of Einstein's GR with two scalar fields, we derive the coefficients $A(r)$ and $C(r)$
that characterize the scalar functions.
We show that the product of these two functions is always negative, which means that either one of the scalars is a ghost.
We use constraints similar to those of mimetic theory to remove such ghosts.
The constraints also make the solution in the model stable under the perturbation.

We assume that the interior of the gravastar is characterized by the de Sitter geometry, which is characterized by the cosmological constant $\lambda$,
and the exterior by the Schwarzschild one, which is characterized by the gravitational mass $M$.
On the shell, we assume a polynomial of fifth order that has six unknown constants (\ref{fr}).
These constants are fixed through the smooth junction condition in the interior geometry and the interior of the shell and
by the smooth junction condition between the exterior geometry and the exterior shell.
Through these fixations, we show that the polynomial function has two maxima when $r<r_1$ and $r>r_2$, respectively, and one minimum when $r_1<r<r_2$,
where $r_1$ and $r_2$ are the radii of the boundaries between the shell region and the interior and exterior regions, respectively.
Through these constraints, we succeeded in writing the two functions $A$ and $C$ and the potential $V$ in terms of $r_1$, $r_2$, and $\lambda$.
We show that on the boundaries of the shell $r=r_1$ and $r=r_2$, the junction conditions are satisfied also for these functions, $A$, $C$, and $V$.

To clarify our philosophy, we give two additional examples: For small $r$ asymptotes of de Sitter spacetime and for large $r$ asymptotes of Schwarzschild geometry.
The first example was given by the geometry similar to the Hayward black hole solution but without horizons.
The second one was the one given by equation (\ref{sad}).
These two examples give the de Sitter for small $r$ and Schwarzschild geometry for large $r$.
We derive the explicative forms of $A$, $C$, and $V$ of these two models and show their behaviors in Figure~\ref{Fig:1}, Figure~\ref{Fig:1B},
and Figure~\ref{Fig:2}, which ensures our analytical derivation.
We calculated the proper length and (surface) redshift of these models as given by Eqs.~(\ref{fr}), (\ref{hayward1}), and (\ref{sad})
and we have shown that there exists a stable gravastar with a large surface redshift prohibited by the instability in the previous works.
Furthermore, we check the behavior of the effective EoS parameters and show that the gravastar geometry could be difficult to be realized by the ordinary fluid.

In conclusion, our study of gravastars within the framework of Einstein's GR, enriched by the inclusion of two scalar fields, offers a promising avenue
for furthering our understanding of exotic astrophysical objects.
We have shown that these scalar fields have a substantial impact on the properties and behavior of gravastars, making them more than just theoretical constructs.

An interesting and very important point is whether the gravastar given in this paper can be formed via gravitational collapse. 
In the paper~\cite{Nojiri:2020blr}, on which this paper is based, it has been shown that for any given arbitrary spherically symmetric spacetime even if the spacetime
is time-dependent, we can construct a model that realizes the given spacetime although we have considered only static spacetime in this paper. 
As discussed in this paper, the model includes ghosts.
The ghosts can be eliminated by imposing the constraints (\ref{lambda2}) because the scalar field $\phi$ is identified with the time coordinate $t$
and the scalar field $\chi$ with the radial coordinate $r$ even in the time-dependent model. 
As a result, the perturbations of these scalar fields from the background do not propagate,  and therefore the time-dependent solution is stable. 
We may then construct a time-dependent model where the spacetime is isotropic and homogeneous in the infinite past $t\to -\infty$ and the spacetime
becomes gravastar given in Section~\ref{S3} in the infinite future $t\to +\infty$ although the model could be complicated a little bit. 
This time dependence could correspond to the gravitational collapse to form the gravastar.
Because the spacetime becomes static in the limit of $t\to \pm \infty$, static isotropic and homogeneous spacetime and static gravastar in this paper are
exact solutions of the model.

Our research not only reaffirms the potential of gravastars as alternatives to black holes but also highlights the versatility and complexity introduced by scalar fields.
The interplay between these fields and the gravastar's structure opens up a wide array of possibilities for future investigations in the fields of astrophysics and theoretical physics.

As we continue to explore the intricacies of gravastars with scalar fields, we look forward to unveiling new layers of understanding in the ever-evolving narrative of our universe.
This study represents a vital contribution to the ongoing quest to unravel the mysteries of compact astrophysical objects and their role in the cosmos.

Is the procedure carried out in this study applicable in the frame of $f(R)$ theory with two scalar fields or within the frame of Gauss-Bonnet theory with two scalar fields?
The answer to these questions will be found elsewhere.

\begin{thebibliography}{63}
\expandafter\ifx\csname natexlab\endcsname\relax\def\natexlab#1{#1}\fi
\expandafter\ifx\csname bibnamefont\endcsname\relax
 \def\bibnamefont#1{#1}\fi
\expandafter\ifx\csname bibfnamefont\endcsname\relax
 \def\bibfnamefont#1{#1}\fi
\expandafter\ifx\csname citenamefont\endcsname\relax
 \def\citenamefont#1{#1}\fi
\expandafter\ifx\csname url\endcsname\relax
 \def\url#1{\texttt{#1}}\fi
\expandafter\ifx\csname urlprefix\endcsname\relax\def\urlprefix{URL }\fi
\providecommand{\bibinfo}[2]{#2}
\providecommand{\eprint}[2][]{\url{#2}}

\bibitem[{\citenamefont{Mitra}(2002)}]{Mitra:2002dh}
\bibinfo{author}{\bibfnamefont{A.}~\bibnamefont{Mitra}},
 \bibinfo{journal}{Found. Phys. Lett.} \textbf{\bibinfo{volume}{15}},
 \bibinfo{pages}{439} (\bibinfo{year}{2002}), \eprint{astro-ph/0207056}.

\bibitem[{\citenamefont{Mazur and Mottola}(2004)}]{Mazur:2004fk}
\bibinfo{author}{\bibfnamefont{P.~O.} \bibnamefont{Mazur}} \bibnamefont{and}
 \bibinfo{author}{\bibfnamefont{E.}~\bibnamefont{Mottola}},
 \bibinfo{journal}{Proc. Nat. Acad. Sci.} \textbf{\bibinfo{volume}{101}},
 \bibinfo{pages}{9545} (\bibinfo{year}{2004}), \eprint{gr-qc/0407075}.

\bibitem[{\citenamefont{Abbott et~al.}(2016)}]{LIGOScientific:2016aoc}
\bibinfo{author}{\bibfnamefont{B.~P.} \bibnamefont{Abbott}}
 \bibnamefont{et~al.} (\bibinfo{collaboration}{LIGO Scientific, Virgo}),
 \bibinfo{journal}{Phys. Rev. Lett.} \textbf{\bibinfo{volume}{116}},
 \bibinfo{pages}{061102} (\bibinfo{year}{2016}), \eprint{1602.03837}.

\bibitem[{\citenamefont{Visser and Wiltshire}(2004)}]{Visser:2003ge}
\bibinfo{author}{\bibfnamefont{M.}~\bibnamefont{Visser}} \bibnamefont{and}
 \bibinfo{author}{\bibfnamefont{D.~L.} \bibnamefont{Wiltshire}},
 \bibinfo{journal}{Class. Quant. Grav.} \textbf{\bibinfo{volume}{21}},
 \bibinfo{pages}{1135} (\bibinfo{year}{2004}), \eprint{gr-qc/0310107}.

\bibitem[{\citenamefont{Cattoen et~al.}(2005)\citenamefont{Cattoen, Faber, and
 Visser}}]{Cattoen:2005he}
\bibinfo{author}{\bibfnamefont{C.}~\bibnamefont{Cattoen}},
 \bibinfo{author}{\bibfnamefont{T.}~\bibnamefont{Faber}}, \bibnamefont{and}
 \bibinfo{author}{\bibfnamefont{M.}~\bibnamefont{Visser}},
 \bibinfo{journal}{Class. Quant. Grav.} \textbf{\bibinfo{volume}{22}},
 \bibinfo{pages}{4189} (\bibinfo{year}{2005}), \eprint{gr-qc/0505137}.

\bibitem[{\citenamefont{Carter}(2005)}]{Carter:2005pi}
\bibinfo{author}{\bibfnamefont{B.~M.~N.} \bibnamefont{Carter}},
 \bibinfo{journal}{Class. Quant. Grav.} \textbf{\bibinfo{volume}{22}},
 \bibinfo{pages}{4551} (\bibinfo{year}{2005}), \eprint{gr-qc/0509087}.

\bibitem[{\citenamefont{Bilic et~al.}(2006)\citenamefont{Bilic, Tupper, and
 Viollier}}]{Bilic:2005sn}
\bibinfo{author}{\bibfnamefont{N.}~\bibnamefont{Bilic}},
 \bibinfo{author}{\bibfnamefont{G.~B.} \bibnamefont{Tupper}},
 \bibnamefont{and} \bibinfo{author}{\bibfnamefont{R.~D.}
 \bibnamefont{Viollier}}, \bibinfo{journal}{JCAP}
 \textbf{\bibinfo{volume}{02}}, \bibinfo{pages}{013} (\bibinfo{year}{2006}),
 \eprint{astro-ph/0503427}.

\bibitem[{\citenamefont{Lobo}(2006)}]{Lobo:2005uf}
\bibinfo{author}{\bibfnamefont{F.~S.~N.} \bibnamefont{Lobo}},
 \bibinfo{journal}{Class. Quant. Grav.} \textbf{\bibinfo{volume}{23}},
 \bibinfo{pages}{1525} (\bibinfo{year}{2006}), \eprint{gr-qc/0508115}.

\bibitem[{\citenamefont{DeBenedictis et~al.}(2006)\citenamefont{DeBenedictis,
 Horvat, Ilijic, Kloster, and Viswanathan}}]{DeBenedictis:2005vp}
\bibinfo{author}{\bibfnamefont{A.}~\bibnamefont{DeBenedictis}},
 \bibinfo{author}{\bibfnamefont{D.}~\bibnamefont{Horvat}},
 \bibinfo{author}{\bibfnamefont{S.}~\bibnamefont{Ilijic}},
 \bibinfo{author}{\bibfnamefont{S.}~\bibnamefont{Kloster}}, \bibnamefont{and}
 \bibinfo{author}{\bibfnamefont{K.~S.} \bibnamefont{Viswanathan}},
 \bibinfo{journal}{Class. Quant. Grav.} \textbf{\bibinfo{volume}{23}},
 \bibinfo{pages}{2303} (\bibinfo{year}{2006}), \eprint{gr-qc/0511097}.

\bibitem[{\citenamefont{Lobo and Arellano}(2007)}]{Lobo:2006xt}
\bibinfo{author}{\bibfnamefont{F.~S.~N.} \bibnamefont{Lobo}} \bibnamefont{and}
 \bibinfo{author}{\bibfnamefont{A.~V.~B.} \bibnamefont{Arellano}},
 \bibinfo{journal}{Class. Quant. Grav.} \textbf{\bibinfo{volume}{24}},
 \bibinfo{pages}{1069} (\bibinfo{year}{2007}), \eprint{gr-qc/0611083}.

\bibitem[{\citenamefont{Horvat and Ilijic}(2007)}]{Horvat:2007qa}
\bibinfo{author}{\bibfnamefont{D.}~\bibnamefont{Horvat}} \bibnamefont{and}
 \bibinfo{author}{\bibfnamefont{S.}~\bibnamefont{Ilijic}},
 \bibinfo{journal}{Class. Quant. Grav.} \textbf{\bibinfo{volume}{24}},
 \bibinfo{pages}{5637} (\bibinfo{year}{2007}), \eprint{0707.1636}.

\bibitem[{\citenamefont{Chirenti and Rezzolla}(2007)}]{Chirenti:2007mk}
\bibinfo{author}{\bibfnamefont{C.~B. M.~H.} \bibnamefont{Chirenti}}
 \bibnamefont{and} \bibinfo{author}{\bibfnamefont{L.}~\bibnamefont{Rezzolla}},
 \bibinfo{journal}{Class. Quant. Grav.} \textbf{\bibinfo{volume}{24}},
 \bibinfo{pages}{4191} (\bibinfo{year}{2007}), \eprint{0706.1513}.

\bibitem[{\citenamefont{Rocha et~al.}(2008)\citenamefont{Rocha, Chan, da~Silva,
 and Wang}}]{Rocha:2008hi}
\bibinfo{author}{\bibfnamefont{P.}~\bibnamefont{Rocha}},
 \bibinfo{author}{\bibfnamefont{R.}~\bibnamefont{Chan}},
 \bibinfo{author}{\bibfnamefont{M.~F.~A.} \bibnamefont{da~Silva}},
 \bibnamefont{and} \bibinfo{author}{\bibfnamefont{A.}~\bibnamefont{Wang}},
 \bibinfo{journal}{JCAP} \textbf{\bibinfo{volume}{11}}, \bibinfo{pages}{010}
 (\bibinfo{year}{2008}), \eprint{0809.4879}.

\bibitem[{\citenamefont{Horvat et~al.}(2009)\citenamefont{Horvat, Ilijic, and
 Marunovic}}]{Horvat:2008ch}
\bibinfo{author}{\bibfnamefont{D.}~\bibnamefont{Horvat}},
 \bibinfo{author}{\bibfnamefont{S.}~\bibnamefont{Ilijic}}, \bibnamefont{and}
 \bibinfo{author}{\bibfnamefont{A.}~\bibnamefont{Marunovic}},
 \bibinfo{journal}{Class. Quant. Grav.} \textbf{\bibinfo{volume}{26}},
 \bibinfo{pages}{025003} (\bibinfo{year}{2009}), \eprint{0807.2051}.

\bibitem[{\citenamefont{Usmani et~al.}(2008)\citenamefont{Usmani, Ghosh,
 Mukhopadhyay, Ray, and Ray}}]{Usmani:2008ce}
\bibinfo{author}{\bibfnamefont{A.~A.} \bibnamefont{Usmani}},
 \bibinfo{author}{\bibfnamefont{P.~P.} \bibnamefont{Ghosh}},
 \bibinfo{author}{\bibfnamefont{U.}~\bibnamefont{Mukhopadhyay}},
 \bibinfo{author}{\bibfnamefont{P.~C.} \bibnamefont{Ray}}, \bibnamefont{and}
 \bibinfo{author}{\bibfnamefont{S.}~\bibnamefont{Ray}}, \bibinfo{journal}{Mon.
 Not. Roy. Astron. Soc.} \textbf{\bibinfo{volume}{386}}, \bibinfo{pages}{L92}
 (\bibinfo{year}{2008}), \eprint{0801.4529}.

\bibitem[{\citenamefont{Turimov et~al.}(2009)\citenamefont{Turimov, Ahmedov,
 and Abdujabbarov}}]{Turimov:2009mu}
\bibinfo{author}{\bibfnamefont{B.~V.} \bibnamefont{Turimov}},
 \bibinfo{author}{\bibfnamefont{B.~J.} \bibnamefont{Ahmedov}},
 \bibnamefont{and} \bibinfo{author}{\bibfnamefont{A.~A.}
 \bibnamefont{Abdujabbarov}}, \bibinfo{journal}{Mod. Phys. Lett. A}
 \textbf{\bibinfo{volume}{24}}, \bibinfo{pages}{733} (\bibinfo{year}{2009}),
 \eprint{0902.0217}.

\bibitem[{\citenamefont{Nandi et~al.}(2009)\citenamefont{Nandi, Zhang, Cai, and
 Panchenko}}]{Nandi:2008ij}
\bibinfo{author}{\bibfnamefont{K.~K.} \bibnamefont{Nandi}},
 \bibinfo{author}{\bibfnamefont{Y.~Z.} \bibnamefont{Zhang}},
 \bibinfo{author}{\bibfnamefont{R.~G.} \bibnamefont{Cai}}, \bibnamefont{and}
 \bibinfo{author}{\bibfnamefont{A.}~\bibnamefont{Panchenko}},
 \bibinfo{journal}{Phys. Rev. D} \textbf{\bibinfo{volume}{79}},
 \bibinfo{pages}{024011} (\bibinfo{year}{2009}), \eprint{0809.4143}.

\bibitem[{\citenamefont{Harko et~al.}(2009)\citenamefont{Harko, Kovacs, and
 Lobo}}]{Harko:2009gc}
\bibinfo{author}{\bibfnamefont{T.}~\bibnamefont{Harko}},
 \bibinfo{author}{\bibfnamefont{Z.}~\bibnamefont{Kovacs}}, \bibnamefont{and}
 \bibinfo{author}{\bibfnamefont{F.~S.~N.} \bibnamefont{Lobo}},
 \bibinfo{journal}{Class. Quant. Grav.} \textbf{\bibinfo{volume}{26}},
 \bibinfo{pages}{215006} (\bibinfo{year}{2009}), \eprint{0905.1355}.

\bibitem[{\citenamefont{Usmani et~al.}(2011)\citenamefont{Usmani, Rahaman,
 Rakib, Ray, Nandi, Kuhfittig, and Hasan}}]{Usmani:2010ac}
\bibinfo{author}{\bibfnamefont{A.~A.} \bibnamefont{Usmani}},
 \bibinfo{author}{\bibfnamefont{F.}~\bibnamefont{Rahaman}},
 \bibinfo{author}{\bibfnamefont{S.~A.} \bibnamefont{Rakib}},
 \bibinfo{author}{\bibfnamefont{S.}~\bibnamefont{Ray}},
 \bibinfo{author}{\bibfnamefont{K.~K.} \bibnamefont{Nandi}},
 \bibinfo{author}{\bibfnamefont{P.~K.~F.} \bibnamefont{Kuhfittig}},
 \bibnamefont{and} \bibinfo{author}{\bibfnamefont{Z.}~\bibnamefont{Hasan}},
 \bibinfo{journal}{Phys. Lett. B} \textbf{\bibinfo{volume}{701}},
 \bibinfo{pages}{388} (\bibinfo{year}{2011}), \eprint{1012.5605}.

\bibitem[{\citenamefont{Rahaman
 et~al.}(2012{\natexlab{a}})\citenamefont{Rahaman, Ray, Usmani, and
 Islam}}]{Rahaman:2011we}
\bibinfo{author}{\bibfnamefont{F.}~\bibnamefont{Rahaman}},
 \bibinfo{author}{\bibfnamefont{S.}~\bibnamefont{Ray}},
 \bibinfo{author}{\bibfnamefont{A.~A.} \bibnamefont{Usmani}},
 \bibnamefont{and} \bibinfo{author}{\bibfnamefont{S.}~\bibnamefont{Islam}},
 \bibinfo{journal}{Phys. Lett. B} \textbf{\bibinfo{volume}{707}},
 \bibinfo{pages}{319} (\bibinfo{year}{2012}{\natexlab{a}}),
 \eprint{1108.4824}.

\bibitem[{\citenamefont{Rahaman
 et~al.}(2012{\natexlab{b}})\citenamefont{Rahaman, Usmani, Ray, and
 Islam}}]{Rahaman:2012wc}
\bibinfo{author}{\bibfnamefont{F.}~\bibnamefont{Rahaman}},
 \bibinfo{author}{\bibfnamefont{A.~A.} \bibnamefont{Usmani}},
 \bibinfo{author}{\bibfnamefont{S.}~\bibnamefont{Ray}}, \bibnamefont{and}
 \bibinfo{author}{\bibfnamefont{S.}~\bibnamefont{Islam}},
 \bibinfo{journal}{Phys. Lett. B} \textbf{\bibinfo{volume}{717}},
 \bibinfo{pages}{1} (\bibinfo{year}{2012}{\natexlab{b}}), \eprint{1205.6796}.

\bibitem[{\citenamefont{Bhar}(2014)}]{Bhar:2014vra}
\bibinfo{author}{\bibfnamefont{P.}~\bibnamefont{Bhar}},
 \bibinfo{journal}{Astrophys. Space Sci.} \textbf{\bibinfo{volume}{354}},
 \bibinfo{pages}{2109} (\bibinfo{year}{2014}).

\bibitem[{\citenamefont{Rahaman et~al.}(2015)\citenamefont{Rahaman,
 Chakraborty, Ray, Usmani, and Islam}}]{Rahaman:2012xx}
\bibinfo{author}{\bibfnamefont{F.}~\bibnamefont{Rahaman}},
 \bibinfo{author}{\bibfnamefont{S.}~\bibnamefont{Chakraborty}},
 \bibinfo{author}{\bibfnamefont{S.}~\bibnamefont{Ray}},
 \bibinfo{author}{\bibfnamefont{A.~A.} \bibnamefont{Usmani}},
 \bibnamefont{and} \bibinfo{author}{\bibfnamefont{S.}~\bibnamefont{Islam}},
 \bibinfo{journal}{Int. J. Theor. Phys.} \textbf{\bibinfo{volume}{54}},
 \bibinfo{pages}{50} (\bibinfo{year}{2015}), \eprint{1209.6291}.

\bibitem[{\citenamefont{Ghosh et~al.}(2017)\citenamefont{Ghosh, Rahaman, Guha,
 and Ray}}]{Ghosh:2015ohi}
\bibinfo{author}{\bibfnamefont{S.}~\bibnamefont{Ghosh}},
 \bibinfo{author}{\bibfnamefont{F.}~\bibnamefont{Rahaman}},
 \bibinfo{author}{\bibfnamefont{B.~K.} \bibnamefont{Guha}}, \bibnamefont{and}
 \bibinfo{author}{\bibfnamefont{S.}~\bibnamefont{Ray}},
 \bibinfo{journal}{Phys. Lett. B} \textbf{\bibinfo{volume}{767}},
 \bibinfo{pages}{380} (\bibinfo{year}{2017}), \eprint{1511.05417}.

\bibitem[{\citenamefont{Ghosh et~al.}(2018)\citenamefont{Ghosh, Ray, Rahaman,
 and Guha}}]{Ghosh:2017flc}
\bibinfo{author}{\bibfnamefont{S.}~\bibnamefont{Ghosh}},
 \bibinfo{author}{\bibfnamefont{S.}~\bibnamefont{Ray}},
 \bibinfo{author}{\bibfnamefont{F.}~\bibnamefont{Rahaman}}, \bibnamefont{and}
 \bibinfo{author}{\bibfnamefont{B.~K.} \bibnamefont{Guha}},
 \bibinfo{journal}{Annals Phys.} \textbf{\bibinfo{volume}{394}},
 \bibinfo{pages}{230} (\bibinfo{year}{2018}), \eprint{1701.01046}.

\bibitem[{\citenamefont{Ghosh et~al.}(2019{\natexlab{a}})\citenamefont{Ghosh,
 Biswas, Rahaman, Guha, and Ray}}]{Ghosh:2019nsi}
\bibinfo{author}{\bibfnamefont{S.}~\bibnamefont{Ghosh}},
 \bibinfo{author}{\bibfnamefont{S.}~\bibnamefont{Biswas}},
 \bibinfo{author}{\bibfnamefont{F.}~\bibnamefont{Rahaman}},
 \bibinfo{author}{\bibfnamefont{B.~K.} \bibnamefont{Guha}}, \bibnamefont{and}
 \bibinfo{author}{\bibfnamefont{S.}~\bibnamefont{Ray}},
 \bibinfo{journal}{Annals Phys.} \textbf{\bibinfo{volume}{411}},
 \bibinfo{pages}{167968} (\bibinfo{year}{2019}{\natexlab{a}}).

\bibitem[{\citenamefont{Ghosh et~al.}(2019{\natexlab{b}})\citenamefont{Ghosh,
 Shee, Ray, Rahaman, and Guha}}]{Ghosh:2019upa}
\bibinfo{author}{\bibfnamefont{S.}~\bibnamefont{Ghosh}},
 \bibinfo{author}{\bibfnamefont{D.}~\bibnamefont{Shee}},
 \bibinfo{author}{\bibfnamefont{S.}~\bibnamefont{Ray}},
 \bibinfo{author}{\bibfnamefont{F.}~\bibnamefont{Rahaman}}, \bibnamefont{and}
 \bibinfo{author}{\bibfnamefont{B.~K.} \bibnamefont{Guha}},
 \bibinfo{journal}{Results Phys.} \textbf{\bibinfo{volume}{14}},
 \bibinfo{pages}{102473} (\bibinfo{year}{2019}{\natexlab{b}}).

\bibitem[{\citenamefont{Chan and da~Silva}(2010)}]{Chan:2010fs}
\bibinfo{author}{\bibfnamefont{R.}~\bibnamefont{Chan}} \bibnamefont{and}
 \bibinfo{author}{\bibfnamefont{M.~F.~A.} \bibnamefont{da~Silva}},
 \bibinfo{journal}{JCAP} \textbf{\bibinfo{volume}{07}}, \bibinfo{pages}{029}
 (\bibinfo{year}{2010}), \eprint{1005.3703}.

\bibitem[{\citenamefont{Chan et~al.}(2011)\citenamefont{Chan, da~Silva,
 da~Rocha, and Wang}}]{Chan:2011wi}
\bibinfo{author}{\bibfnamefont{R.}~\bibnamefont{Chan}},
 \bibinfo{author}{\bibfnamefont{M.~F.~A.} \bibnamefont{da~Silva}},
 \bibinfo{author}{\bibfnamefont{J.~F.~V.} \bibnamefont{da~Rocha}},
 \bibnamefont{and} \bibinfo{author}{\bibfnamefont{A.}~\bibnamefont{Wang}},
 \bibinfo{journal}{JCAP} \textbf{\bibinfo{volume}{10}}, \bibinfo{pages}{013}
 (\bibinfo{year}{2011}), \eprint{1109.2062}.

\bibitem[{\citenamefont{Riess et~al.}(1998)}]{SupernovaSearchTeam:1998fmf}
\bibinfo{author}{\bibfnamefont{A.~G.} \bibnamefont{Riess}} \bibnamefont{et~al.}
 (\bibinfo{collaboration}{Supernova Search Team}), \bibinfo{journal}{Astron.
 J.} \textbf{\bibinfo{volume}{116}}, \bibinfo{pages}{1009}
 (\bibinfo{year}{1998}), \eprint{astro-ph/9805201}.

\bibitem[{\citenamefont{Perlmutter
 et~al.}(1999)}]{SupernovaCosmologyProject:1998vns}
\bibinfo{author}{\bibfnamefont{S.}~\bibnamefont{Perlmutter}}
 \bibnamefont{et~al.} (\bibinfo{collaboration}{Supernova Cosmology Project}),
 \bibinfo{journal}{Astrophys. J.} \textbf{\bibinfo{volume}{517}},
 \bibinfo{pages}{565} (\bibinfo{year}{1999}), \eprint{astro-ph/9812133}.

\bibitem[{\citenamefont{de~Bernardis et~al.}(2000)}]{Boomerang:2000efg}
\bibinfo{author}{\bibfnamefont{P.}~\bibnamefont{de~Bernardis}}
 \bibnamefont{et~al.} (\bibinfo{collaboration}{Boomerang}),
 \bibinfo{journal}{Nature} \textbf{\bibinfo{volume}{404}},
 \bibinfo{pages}{955} (\bibinfo{year}{2000}), \eprint{astro-ph/0004404}.

\bibitem[{\citenamefont{Hanany et~al.}(2000)}]{Hanany:2000qf}
\bibinfo{author}{\bibfnamefont{S.}~\bibnamefont{Hanany}} \bibnamefont{et~al.},
 \bibinfo{journal}{Astrophys. J. Lett.} \textbf{\bibinfo{volume}{545}},
 \bibinfo{pages}{L5} (\bibinfo{year}{2000}), \eprint{astro-ph/0005123}.

\bibitem[{\citenamefont{Peebles and Ratra}(2003)}]{Peebles:2002gy}
\bibinfo{author}{\bibfnamefont{P.~J.~E.} \bibnamefont{Peebles}}
 \bibnamefont{and} \bibinfo{author}{\bibfnamefont{B.}~\bibnamefont{Ratra}},
 \bibinfo{journal}{Rev. Mod. Phys.} \textbf{\bibinfo{volume}{75}},
 \bibinfo{pages}{559} (\bibinfo{year}{2003}), \eprint{astro-ph/0207347}.

\bibitem[{\citenamefont{Padmanabhan}(2003)}]{Padmanabhan:2002ji}
\bibinfo{author}{\bibfnamefont{T.}~\bibnamefont{Padmanabhan}},
 \bibinfo{journal}{Phys. Rept.} \textbf{\bibinfo{volume}{380}},
 \bibinfo{pages}{235} (\bibinfo{year}{2003}), \eprint{hep-th/0212290}.

\bibitem[{\citenamefont{Clifton et~al.}(2012)\citenamefont{Clifton, Ferreira,
 Padilla, and Skordis}}]{Clifton:2011jh}
\bibinfo{author}{\bibfnamefont{T.}~\bibnamefont{Clifton}},
 \bibinfo{author}{\bibfnamefont{P.~G.} \bibnamefont{Ferreira}},
 \bibinfo{author}{\bibfnamefont{A.}~\bibnamefont{Padilla}}, \bibnamefont{and}
 \bibinfo{author}{\bibfnamefont{C.}~\bibnamefont{Skordis}},
 \bibinfo{journal}{Physics reports} \textbf{\bibinfo{volume}{513}},
 \bibinfo{pages}{1} (\bibinfo{year}{2012}).

\bibitem[{\citenamefont{Riess et~al.}(2007)}]{Riess:2006fw}
\bibinfo{author}{\bibfnamefont{A.~G.} \bibnamefont{Riess}}
 \bibnamefont{et~al.}, \bibinfo{journal}{Astrophys. J.}
 \textbf{\bibinfo{volume}{659}}, \bibinfo{pages}{98} (\bibinfo{year}{2007}),
 \eprint{astro-ph/0611572}.

\bibitem[{\citenamefont{Tegmark et~al.}(2004)}]{SDSS:2003eyi}
\bibinfo{author}{\bibfnamefont{M.}~\bibnamefont{Tegmark}} \bibnamefont{et~al.}
 (\bibinfo{collaboration}{SDSS}), \bibinfo{journal}{Phys. Rev. D}
 \textbf{\bibinfo{volume}{69}}, \bibinfo{pages}{103501}
 (\bibinfo{year}{2004}), \eprint{astro-ph/0310723}.

\bibitem[{\citenamefont{Amanullah et~al.}(2010)}]{Amanullah:2010vv}
\bibinfo{author}{\bibfnamefont{R.}~\bibnamefont{Amanullah}}
 \bibnamefont{et~al.}, \bibinfo{journal}{Astrophys. J.}
 \textbf{\bibinfo{volume}{716}}, \bibinfo{pages}{712} (\bibinfo{year}{2010}),
 \eprint{1004.1711}.

\bibitem[{\citenamefont{Komatsu et~al.}(2011)}]{WMAP:2010qai}
\bibinfo{author}{\bibfnamefont{E.}~\bibnamefont{Komatsu}} \bibnamefont{et~al.}
 (\bibinfo{collaboration}{WMAP}), \bibinfo{journal}{Astrophys. J. Suppl.}
 \textbf{\bibinfo{volume}{192}}, \bibinfo{pages}{18} (\bibinfo{year}{2011}),
 \eprint{1001.4538}.

\bibitem[{\citenamefont{Das et~al.}(2015)\citenamefont{Das, Rahaman, Guha, and
 Ray}}]{Das:2015gwa}
\bibinfo{author}{\bibfnamefont{A.}~\bibnamefont{Das}},
 \bibinfo{author}{\bibfnamefont{F.}~\bibnamefont{Rahaman}},
 \bibinfo{author}{\bibfnamefont{B.~K.} \bibnamefont{Guha}}, \bibnamefont{and}
 \bibinfo{author}{\bibfnamefont{S.}~\bibnamefont{Ray}},
 \bibinfo{journal}{Astrophys. Space Sci.} \textbf{\bibinfo{volume}{358}},
 \bibinfo{pages}{36} (\bibinfo{year}{2015}), \eprint{1507.04959}.

\bibitem[{\citenamefont{Das et~al.}(2016)\citenamefont{Das, Rahaman, Guha, and
 Ray}}]{Das:2016mxq}
\bibinfo{author}{\bibfnamefont{A.}~\bibnamefont{Das}},
 \bibinfo{author}{\bibfnamefont{F.}~\bibnamefont{Rahaman}},
 \bibinfo{author}{\bibfnamefont{B.}~\bibnamefont{Guha}}, \bibnamefont{and}
 \bibinfo{author}{\bibfnamefont{S.}~\bibnamefont{Ray}}, \bibinfo{journal}{The
 European Physical Journal C} \textbf{\bibinfo{volume}{76}},
 \bibinfo{pages}{1} (\bibinfo{year}{2016}).

\bibitem[{\citenamefont{Das et~al.}(2017)\citenamefont{Das, Ghosh, Guha, Das,
 Rahaman, and Ray}}]{Das:2017rhi}
\bibinfo{author}{\bibfnamefont{A.}~\bibnamefont{Das}},
 \bibinfo{author}{\bibfnamefont{S.}~\bibnamefont{Ghosh}},
 \bibinfo{author}{\bibfnamefont{B.~K.} \bibnamefont{Guha}},
 \bibinfo{author}{\bibfnamefont{S.}~\bibnamefont{Das}},
 \bibinfo{author}{\bibfnamefont{F.}~\bibnamefont{Rahaman}}, \bibnamefont{and}
 \bibinfo{author}{\bibfnamefont{S.}~\bibnamefont{Ray}},
 \bibinfo{journal}{Phys. Rev. D} \textbf{\bibinfo{volume}{95}},
 \bibinfo{pages}{124011} (\bibinfo{year}{2017}), \eprint{1702.08873}.

\bibitem[{\citenamefont{Biswas et~al.}(2019)\citenamefont{Biswas, Ghosh, Ray,
 Rahaman, and Guha}}]{Biswas:2018inc}
\bibinfo{author}{\bibfnamefont{S.}~\bibnamefont{Biswas}},
 \bibinfo{author}{\bibfnamefont{S.}~\bibnamefont{Ghosh}},
 \bibinfo{author}{\bibfnamefont{S.}~\bibnamefont{Ray}},
 \bibinfo{author}{\bibfnamefont{F.}~\bibnamefont{Rahaman}}, \bibnamefont{and}
 \bibinfo{author}{\bibfnamefont{B.~K.} \bibnamefont{Guha}},
 \bibinfo{journal}{Annals Phys.} \textbf{\bibinfo{volume}{401}},
 \bibinfo{pages}{1} (\bibinfo{year}{2019}), \eprint{1803.00442}.

\bibitem[{\citenamefont{Ghosh et~al.}(2020)\citenamefont{Ghosh, Kanfon, Das,
 Houndjo, Salako, and Ray}}]{Ghosh:2020rau}
\bibinfo{author}{\bibfnamefont{S.}~\bibnamefont{Ghosh}},
 \bibinfo{author}{\bibfnamefont{A.~D.} \bibnamefont{Kanfon}},
 \bibinfo{author}{\bibfnamefont{A.}~\bibnamefont{Das}},
 \bibinfo{author}{\bibfnamefont{M.~J.~S.} \bibnamefont{Houndjo}},
 \bibinfo{author}{\bibfnamefont{I.~G.} \bibnamefont{Salako}},
 \bibnamefont{and} \bibinfo{author}{\bibfnamefont{S.}~\bibnamefont{Ray}},
 \bibinfo{journal}{Int. J. Mod. Phys. A} \textbf{\bibinfo{volume}{35}},
 \bibinfo{pages}{2050017} (\bibinfo{year}{2020}), \eprint{2003.03194}.

\bibitem[{\citenamefont{Das et~al.}(2020)\citenamefont{Das, Ghosh, Deb,
 Rahaman, and Ray}}]{Das:2020lqy}
\bibinfo{author}{\bibfnamefont{A.}~\bibnamefont{Das}},
 \bibinfo{author}{\bibfnamefont{S.}~\bibnamefont{Ghosh}},
 \bibinfo{author}{\bibfnamefont{D.}~\bibnamefont{Deb}},
 \bibinfo{author}{\bibfnamefont{F.}~\bibnamefont{Rahaman}}, \bibnamefont{and}
 \bibinfo{author}{\bibfnamefont{S.}~\bibnamefont{Ray}},
 \bibinfo{journal}{Nucl. Phys. B} \textbf{\bibinfo{volume}{954}},
 \bibinfo{pages}{114986} (\bibinfo{year}{2020}), \eprint{2003.10842}.

\bibitem[{\citenamefont{Sengupta et~al.}(2020)\citenamefont{Sengupta, Ghosh,
 Ray, Mishra, and Tripathy}}]{Sengupta:2020lhw}
\bibinfo{author}{\bibfnamefont{R.}~\bibnamefont{Sengupta}},
 \bibinfo{author}{\bibfnamefont{S.}~\bibnamefont{Ghosh}},
 \bibinfo{author}{\bibfnamefont{S.}~\bibnamefont{Ray}},
 \bibinfo{author}{\bibfnamefont{B.}~\bibnamefont{Mishra}}, \bibnamefont{and}
 \bibinfo{author}{\bibfnamefont{S.~K.} \bibnamefont{Tripathy}},
 \bibinfo{journal}{Phys. Rev. D} \textbf{\bibinfo{volume}{102}},
 \bibinfo{pages}{024037} (\bibinfo{year}{2020}), \eprint{2004.01519}.

\bibitem[{\citenamefont{Banerjee et~al.}(2020)\citenamefont{Banerjee, Ghosh,
 Paul, and Rahaman}}]{Banerjee:2020zjq}
\bibinfo{author}{\bibfnamefont{S.}~\bibnamefont{Banerjee}},
 \bibinfo{author}{\bibfnamefont{S.}~\bibnamefont{Ghosh}},
 \bibinfo{author}{\bibfnamefont{N.}~\bibnamefont{Paul}}, \bibnamefont{and}
 \bibinfo{author}{\bibfnamefont{F.}~\bibnamefont{Rahaman}},
 \bibinfo{journal}{Eur. Phys. J. Plus} \textbf{\bibinfo{volume}{135}},
 \bibinfo{pages}{185} (\bibinfo{year}{2020}).

\bibitem[{\citenamefont{Nojiri et~al.}(2021)\citenamefont{Nojiri, Odintsov, and
 Faraoni}}]{Nojiri:2020blr}
\bibinfo{author}{\bibfnamefont{S.}~\bibnamefont{Nojiri}},
 \bibinfo{author}{\bibfnamefont{S.~D.} \bibnamefont{Odintsov}},
 \bibnamefont{and} \bibinfo{author}{\bibfnamefont{V.}~\bibnamefont{Faraoni}},
 \bibinfo{journal}{Phys. Rev. D} \textbf{\bibinfo{volume}{103}},
 \bibinfo{pages}{044055} (\bibinfo{year}{2021}), \eprint{2010.11790}.

\bibitem[{\citenamefont{Chamseddine and Mukhanov}(2013)}]{Chamseddine:2013kea}
\bibinfo{author}{\bibfnamefont{A.~H.} \bibnamefont{Chamseddine}}
 \bibnamefont{and} \bibinfo{author}{\bibfnamefont{V.}~\bibnamefont{Mukhanov}},
 \bibinfo{journal}{JHEP} \textbf{\bibinfo{volume}{11}}, \bibinfo{pages}{135}
 (\bibinfo{year}{2013}), \eprint{1308.5410}.

\bibitem[{\citenamefont{Chamseddine et~al.}(2014)\citenamefont{Chamseddine,
 Mukhanov, and Vikman}}]{Chamseddine:2014vna}
\bibinfo{author}{\bibfnamefont{A.~H.} \bibnamefont{Chamseddine}},
 \bibinfo{author}{\bibfnamefont{V.}~\bibnamefont{Mukhanov}}, \bibnamefont{and}
 \bibinfo{author}{\bibfnamefont{A.}~\bibnamefont{Vikman}},
 \bibinfo{journal}{JCAP} \textbf{\bibinfo{volume}{06}}, \bibinfo{pages}{017}
 (\bibinfo{year}{2014}), \eprint{1403.3961}.

\bibitem[{\citenamefont{Nojiri and Nashed}(2022)}]{Nojiri:2022cah}
\bibinfo{author}{\bibfnamefont{S.}~\bibnamefont{Nojiri}} \bibnamefont{and}
 \bibinfo{author}{\bibfnamefont{G.~G.~L.} \bibnamefont{Nashed}},
 \bibinfo{journal}{Phys. Lett. B} \textbf{\bibinfo{volume}{830}},
 \bibinfo{pages}{137140} (\bibinfo{year}{2022}), \eprint{2202.03693}.
\bibitem[{\citenamefont{Myrzakulov et~al.}(2016)\citenamefont{Myrzakulov,
 Sebastiani, Vagnozzi, and Zerbini}}]{Myrzakulov:2015kda}
\bibinfo{author}{\bibfnamefont{R.}~\bibnamefont{Myrzakulov}},
 \bibinfo{author}{\bibfnamefont{L.}~\bibnamefont{Sebastiani}},
 \bibinfo{author}{\bibfnamefont{S.}~\bibnamefont{Vagnozzi}}, \bibnamefont{and}
 \bibinfo{author}{\bibfnamefont{S.}~\bibnamefont{Zerbini}},
 \bibinfo{journal}{Class. Quant. Grav.} \textbf{\bibinfo{volume}{33}},
 \bibinfo{pages}{125005} (\bibinfo{year}{2016}), \eprint{1510.02284}.

\bibitem[{\citenamefont{Nashed and Nojiri}(2022)}]{Nashed:2021hgn}
\bibinfo{author}{\bibfnamefont{G.~G.~L.} \bibnamefont{Nashed}}
 \bibnamefont{and} \bibinfo{author}{\bibfnamefont{S.}~\bibnamefont{Nojiri}},
 \bibinfo{journal}{JCAP} \textbf{\bibinfo{volume}{05}}, \bibinfo{pages}{011}
 (\bibinfo{year}{2022}), \eprint{2110.08560}.

\bibitem[{\citenamefont{Nashed}(2023{\natexlab{a}})}]{Nashed:2023jdf}
\bibinfo{author}{\bibfnamefont{G.~G.~L.} \bibnamefont{Nashed}},
 \bibinfo{journal}{Nucl. Phys. B} \textbf{\bibinfo{volume}{993}},
 \bibinfo{pages}{116264} (\bibinfo{year}{2023}{\natexlab{a}}),
 \eprint{2307.03199}.

\bibitem[{\citenamefont{Nashed}(2023{\natexlab{b}})}]{Nashed:2023fzp}
\bibinfo{author}{\bibfnamefont{G.~G.~L.} \bibnamefont{Nashed}},
 \bibinfo{journal}{Gen. Rel. Grav.} \textbf{\bibinfo{volume}{55}},
 \bibinfo{pages}{63} (\bibinfo{year}{2023}{\natexlab{b}}),
 \eprint{2305.09694}.

 \bibitem[{\citenamefont{Myrzakulov et~al.}(2015)\citenamefont{Myrzakulov,
 Sebastiani, and Vagnozzi}}]{Myrzakulov:2015qaa}
\bibinfo{author}{\bibfnamefont{R.}~\bibnamefont{Myrzakulov}},
 \bibinfo{author}{\bibfnamefont{L.}~\bibnamefont{Sebastiani}},
 \bibnamefont{and} \bibinfo{author}{\bibfnamefont{S.}~\bibnamefont{Vagnozzi}},
 \bibinfo{journal}{Eur. Phys. J. C} \textbf{\bibinfo{volume}{75}},
 \bibinfo{pages}{444} (\bibinfo{year}{2015}), \eprint{1504.07984}.
\bibitem[{\citenamefont{Vagnozzi}(2017)}]{Vagnozzi:2017ilo}
\bibinfo{author}{\bibfnamefont{S.}~\bibnamefont{Vagnozzi}},
 \bibinfo{journal}{Class. Quant. Grav.} \textbf{\bibinfo{volume}{34}},
 \bibinfo{pages}{185006} (\bibinfo{year}{2017}), \eprint{1708.00603}.

\bibitem[{\citenamefont{Nashed and Nojiri}(2021)}]{Nashed:2021ctg}
\bibinfo{author}{\bibfnamefont{G.~G.~L.} \bibnamefont{Nashed}}
 \bibnamefont{and} \bibinfo{author}{\bibfnamefont{S.}~\bibnamefont{Nojiri}},
 \bibinfo{journal}{Phys. Rev. D} \textbf{\bibinfo{volume}{104}},
 \bibinfo{pages}{044043} (\bibinfo{year}{2021}), \eprint{2107.13550}.

 \bibitem[{\citenamefont{Casalino et~al.}(2018)\citenamefont{Casalino, Rinaldi,
 Sebastiani, and Vagnozzi}}]{Casalino:2018tcd}
\bibinfo{author}{\bibfnamefont{A.}~\bibnamefont{Casalino}},
 \bibinfo{author}{\bibfnamefont{M.}~\bibnamefont{Rinaldi}},
 \bibinfo{author}{\bibfnamefont{L.}~\bibnamefont{Sebastiani}},
 \bibnamefont{and} \bibinfo{author}{\bibfnamefont{S.}~\bibnamefont{Vagnozzi}},
 \bibinfo{journal}{Phys. Dark Univ.} \textbf{\bibinfo{volume}{22}},
 \bibinfo{pages}{108} (\bibinfo{year}{2018}), \eprint{1803.02620}.

\bibitem[{\citenamefont{Casalino et~al.}(2019)\citenamefont{Casalino, Rinaldi,
 Sebastiani, and Vagnozzi}}]{Casalino:2018wnc}
\bibinfo{author}{\bibfnamefont{A.}~\bibnamefont{Casalino}},
 \bibinfo{author}{\bibfnamefont{M.}~\bibnamefont{Rinaldi}},
 \bibinfo{author}{\bibfnamefont{L.}~\bibnamefont{Sebastiani}},
 \bibnamefont{and} \bibinfo{author}{\bibfnamefont{S.}~\bibnamefont{Vagnozzi}},
 \bibinfo{journal}{Class. Quant. Grav.} \textbf{\bibinfo{volume}{36}},
 \bibinfo{pages}{017001} (\bibinfo{year}{2019}), \eprint{1811.06830}.

 \bibitem[{\citenamefont{Sebastiani et~al.}(2017)\citenamefont{Sebastiani,
 Vagnozzi, and Myrzakulov}}]{Sebastiani:2016ras}
\bibinfo{author}{\bibfnamefont{L.}~\bibnamefont{Sebastiani}},
 \bibinfo{author}{\bibfnamefont{S.}~\bibnamefont{Vagnozzi}}, \bibnamefont{and}
 \bibinfo{author}{\bibfnamefont{R.}~\bibnamefont{Myrzakulov}},
 \bibinfo{journal}{Adv. High Energy Phys.} \textbf{\bibinfo{volume}{2017}},
 \bibinfo{pages}{3156915} (\bibinfo{year}{2017}), \eprint{1612.08661}.

\bibitem[{\citenamefont{Nashed}(2021)}]{Nashed:2021pkc}
\bibinfo{author}{\bibfnamefont{G.~G.~L.} \bibnamefont{Nashed}},
 \bibinfo{journal}{Astrophys. J.} \textbf{\bibinfo{volume}{919}},
 \bibinfo{pages}{113} (\bibinfo{year}{2021}), \eprint{2108.04060}.

\bibitem{EventHorizonTelescope:2019ths}
K.~Akiyama \textit{et al.} [Event Horizon Telescope],
Astrophys. J. Lett. \textbf{875} (2019) no.1, L4
doi:10.3847/2041-8213/ab0e85
[arXiv:1906.11241 [astro-ph.GA]].

\bibitem[{\citenamefont{Hayward}(2006)}]{Hayward:2005gi}
\bibinfo{author}{\bibfnamefont{S.~A.} \bibnamefont{Hayward}},
 \bibinfo{journal}{Phys. Rev. Lett.} \textbf{\bibinfo{volume}{96}},
 \bibinfo{pages}{031103} (\bibinfo{year}{2006}), \eprint{gr-qc/0506126}.

\bibitem[{\citenamefont{Buchdahl}(1959)}]{Buchdahl:1959zz}
\bibinfo{author}{\bibfnamefont{H.~A.} \bibnamefont{Buchdahl}},
 \bibinfo{journal}{Physical Review} \textbf{\bibinfo{volume}{116}},
 \bibinfo{pages}{1027} (\bibinfo{year}{1959}).

\bibitem[{\citenamefont{Straumann}(1984)}]{Straumann:1984xf}
\bibinfo{author}{\bibfnamefont{N.}~\bibnamefont{Straumann}},
 \emph{\bibinfo{title}{{GENERAL RELATIVITY AND RELATIVISTIC ASTROPHYSICS}}}
 (\bibinfo{year}{1984}).

\bibitem[{\citenamefont{B{\"o}hmer and Harko}(2006)}]{Boehmer:2006ye}
\bibinfo{author}{\bibfnamefont{C.}~\bibnamefont{B{\"o}hmer}} \bibnamefont{and}
 \bibinfo{author}{\bibfnamefont{T.}~\bibnamefont{Harko}},
 \bibinfo{journal}{Classical and Quantum Gravity}
 \textbf{\bibinfo{volume}{23}}, \bibinfo{pages}{6479} (\bibinfo{year}{2006}).

\bibitem[{\citenamefont{Ivanov}(2002)}]{Ivanov:2002xf}
\bibinfo{author}{\bibfnamefont{B.~V.} \bibnamefont{Ivanov}},
 \bibinfo{journal}{Physical Review D} \textbf{\bibinfo{volume}{65}},
 \bibinfo{pages}{104011} (\bibinfo{year}{2002}).

\bibitem[{\citenamefont{Barraco and Hamity}(2002)}]{Barraco:2002ds}
\bibinfo{author}{\bibfnamefont{D.}~\bibnamefont{Barraco}} \bibnamefont{and}
 \bibinfo{author}{\bibfnamefont{V.~H.} \bibnamefont{Hamity}},
 \bibinfo{journal}{Phys. Rev. D} \textbf{\bibinfo{volume}{65}},
 \bibinfo{pages}{124028} (\bibinfo{year}{2002}).

\end{thebibliography}


\end{document}